\documentclass[twocolumn]{aastex63}

\received{August 20, 2021}
\accepted{September 9, 2021}

\shorttitle{Ionized Calcium in WASP-76b}
\shortauthors{Deibert et al.}

\graphicspath{{./}{figures/}}

\usepackage{printlen}

\begin{document}

\title{Detection of Ionized Calcium in the Atmosphere of the Ultra-Hot Jupiter WASP-76b}

\correspondingauthor{Emily K. Deibert}
\email{deibert@astro.utoronto.ca}

\author[0000-0001-9796-2158]{Emily K. Deibert}
\affiliation{David A. Dunlap Department of Astronomy \& Astrophysics, University of Toronto, Toronto, ON M5S 3H4, Canada}
\affiliation{Dunlap Institute for Astronomy \& Astrophysics, University of Toronto, Toronto, ON M5S 3H4, Canada}

\author[0000-0001-6391-9266]{Ernst J. W. de Mooij}
\affil{Astrophysics Research Centre, Queen's University Belfast, Belfast BT7 1NN, UK}

\author[0000-0001-5349-6853]{Ray Jayawardhana}
\affil{Department of Astronomy, Cornell University, Ithaca, New York 14853, USA}

\author[0000-0001-7836-1787]{Jake D. Turner}
\affil{Department of Astronomy, Cornell University, Ithaca, New York 14853, USA}

\author[0000-0002-5425-2655]{Andrew Ridden-Harper}
\affil{Department of Astronomy, Cornell University, Ithaca, New York 14853, USA}

\author[0000-0003-4426-9530]{Luca Fossati}
\affil{Space Research Institute, Austrian Academy of Sciences, Schmiedlstrasse 6, A-8042 Graz, Austria}

\author[0000-0003-1150-7889]{Callie E. Hood}
\affil{Department of Astronomy \& Astrophysics, University of California, Santa Cruz, CA 95064, USA}

\author[0000-0002-9843-4354]{Jonathan J. Fortney}
\affil{Department of Astronomy \& Astrophysics, University of California, Santa Cruz, CA 95064, USA}

\author[0000-0001-6362-0571]{Laura Flagg}
\affil{Department of Astronomy, Cornell University, Ithaca, New York 14853, USA}

\author[0000-0003-4816-3469]{Ryan MacDonald}
\affil{Department of Astronomy, Cornell University, Ithaca, New York 14853, USA}

\author[0000-0002-1199-9759]{Romain Allart}
\affil{Department of Physics, and Institute for Research on Exoplanets, Universit\'e de Montr\'eal, Montr\'eal, H3T 1J4, Canada}

\author[0000-0001-6050-7645]{David K. Sing}
\affil{Department of Earth \& Planetary Sciences, Johns Hopkins University, Baltimore, MD, USA}
\affil{Department of Physics \& Astronomy, Johns Hopkins University, Baltimore, MD, USA}

\begin{abstract}
Recent observations of the ultra-hot Jupiter WASP-76b have revealed a diversity of atmospheric species. Here we present new high-resolution transit spectroscopy of WASP-76b with GRACES at the Gemini North Observatory, serving as a baseline for the Large and Long Program ``Exploring the Diversity of Exoplanet Atmospheres at High Spectral Resolution'' (Exoplanets with Gemini Spectroscopy, or ExoGemS for short). With a broad spectral range of 400 -- 1050 nm, these observations allow us to search for a suite of atomic species.
We recover absorption features due to neutral sodium (Na~I), and report a new detection of the ionized calcium (Ca~II) triplet at $\sim$ 850 nm in the atmosphere of WASP-76b, complementing a previous detection of the Ca~II H \& K lines. The triplet has line depths of $0.295 \pm 0.034$\% at $\sim$ 849.2 nm, $0.574 \pm 0.041$\% at $\sim$ 854.2 nm, and $0.454 \pm 0.024$\% at $\sim$ 866.2 nm, corresponding to effective radii close to (but within) the planet's Roche radius. These measured line depths are significantly larger than those predicted by model LTE and NLTE spectra obtained on the basis of a pressure-temperature profile computed assuming radiative equilibrium. The discrepancy suggests that the layers probed by our observations are either significantly hotter than predicted by radiative equilibrium and/or in a hydrodynamic state. Our results shed light on the exotic atmosphere of this ultra-hot world, and will inform future analyses from the ExoGemS survey.
\end{abstract}


\section{Introduction}
\label{sec:intro}
In recent years, high-resolution transmission and emission spectroscopy with ground-based observatories has revealed numerous atomic \citep[e.g.,][]{Snellen08, Seidel19, Turner20} and molecular \citep[e.g.,][]{Snellen10,Hawker18} species in exoplanet atmospheres. Such observations not only provide us with insight into their chemical compositions, but also allow us to investigate winds \citep[e.g.,][]{Snellen10, Seidel21}, planetary rotation patterns \citep[e.g.,][]{Snellen14, Brogi16}, and atmospheric escape \citep[e.g.,][]{Nortmann18, Allart18}, shedding light on the underlying physical processes in these alien worlds.
Following numerous detections for individual exoplanets, it will soon be possible to compare trends across populations and move into the field of comparative exoplanetology at high resolution (as has been done for low-resolution spectra; e.g., \citealt{Sing2016}).

Several studies have established the potential of GRACES (the Gemini Remote Access to CFHT ESPaDOnS Spectrograph; \citealt{GRACES}) at the Gemini North Telescope in characterizing atmospheres of planets ranging from super-Earths \citep[e.g.,][]{Esteves17} to gas giants  \citep[e.g.,][]{Deibert19}. The Gemini Large and Long Program GN-2020B-LP106: ``Exploring the Diversity of Exoplanet Atmospheres at High Spectral Resolution'' (Exoplanets with Gemini Spectroscopy or ExoGemS for short; PI: Jake Turner) aims to harness this potential by carrying out the first systematic, high-resolution, comparative survey of transiting exoplanets with GRACES. With planned observations of more than 40 exoplanets, the survey will probe
atmospheres over a range of planetary masses and temperatures, illuminating the role of these parameters in regulating atmospheres and yielding insights into planetary formation and evolution.

Serving as a benchmark target for ExoGemS, WASP-76b \citep{West16} is a transiting ultra-hot Jupiter (i.e., with $T_{\text{eq}} \gtrsim$ 2200 K, \citealt{Bell18, Parmentier18, Arcangeli18}) that orbits its bright ($V$ = 9.52; \citealt{SIMBAD}) F7 V host star in $\sim$ 1.8 days \citep{Ehrenreich20}.
With a mass of $\sim$ 0.89 $M_{\mathrm{J}}$ (recently refined with radial velocity measurements; \citealt{Ehrenreich20}) and an inflated radius of $\sim$ 1.85 $R_{\mathrm{J}}$ \citep{Ehrenreich20}, the planet is ideally suited for atmospheric characterization and has received considerable attention. 

Several atomic and ionic species have been detected recently in its atmosphere. Using HARPS (the High-Accuracy Radial-velocity Planet Searcher; \citealt{HARPS}), 
\cite{Seidel19} reported a detection of the Na~I doublet near 589 nm. Data from the ESPRESSO (Echelle Spectrograph for Rocky Exoplanets and Stable Spectroscopic Observations; \citealt{ESPRESSO}) instrument have provided additional insights  \citep[][]{Ehrenreich20, Tabernero20, Seidel21}. Using two transits observed with ESPRESSO, \cite{Ehrenreich20} reported asymmetric absorption due to neutral iron; their result was confirmed by \cite{Kesseli21} using archival HARPS data of four transits, and later investigated through 3D Monte-Carlo radiative transfer modelling by \cite{Wardenier21}. \cite{Tabernero20} carried out a detailed analysis of the spectra presented in \cite{Ehrenreich20}, searching for absorption due to a range of atomic and molecular species. They confirmed the sodium detection of \cite{Seidel19} and the iron detection of \cite{Ehrenreich20}, and reported detections of Li I, Mg I, Ca~II, Mn I, and K~I. They did not detect, but provided upper limits for, Ti I, Cr I, Ni I, TiO, VO, and ZrO. 

In this Letter, we present, for the first time in WASP-76b's atmosphere, the detection of the ionized calcium (Ca~II) triplet at $\sim$ 850 nm, using transit observations with GRACES\footnote{We note that after this Letter was referreed, \cite{CB21} reported a detection of the Ca~II triplet in the atmosphere of WASP-76b using transit observations from CARMENES/Calar Alto. Our results agree with the results presented therein.}.
We also confirm the Na~I doublet \citep[][]{Seidel19,Tabernero20}, tentatively recover Li~I and K~I signals \citep[][]{Tabernero20}, and report a tentative non-detection of H$\alpha$. This work represents the first results from the ExoGemS survey, and the fifth detection of the Ca~II triplet in ultra-hot Jupiters.

In Section \ref{sec:obs}, we describe our high-resolution optical spectra obtained using GRACES. Our data reduction process, including the removal of telluric and stellar absorption features and corrections for the centre-to-limb variation (CLV) and the Rossiter-McLaughlin (RM) effect, is presented in Section \ref{sec:reduc}. Our analysis and modelling routines are described in Section \ref{sec:analysis}, while the results and discussion follow in Section \ref{sec:discussion}, with a conclusion in Section \ref{sec:conc}. The Appendices present additional details. 

\section{Observations}
\label{sec:obs}
We observed one transit of WASP-76b with GRACES, which uses a fibre optic feed to combine the large collecting area of the Gemini North Telescope with the high resolving power of the ESPaDOnS (Echelle SpectroPolarimetric Device for the Observation of Stars; \citealt{espadons}) spectrograph at the Canada France Hawaii Telescope (CFHT). GRACES spectra cover the optical wavelengths from 400 to 1050 nm at a resolving power of R $\sim$ 66,000. 
In total, 169 spectra (120 in-transit and 49 out-of-transit) were obtained over a period of $\sim$ 5.16 hours on 11 October 2020, including observations before, during, and after transit. A full summary of the observing conditions can be found in Appendix \ref{app:obscond}.

\begin{deluxetable*}{lccc}
\label{tab:parameters}
\tablecaption{Orbital and physical parameters of the WASP-76 system used in this analysis.}
\tablehead{%
    \colhead{Parameter} & \colhead{Symbol (Unit)} & \colhead{Value} & \colhead{Reference}
    }
\startdata
Stellar radius & $R_*$ ($R_\sun$) & $1.756 \pm 0.071$ & \cite{Ehrenreich20} \\
Stellar mass & $M_*$ ($M_\sun$) & $1.458 \pm 0.02$ & \cite{Ehrenreich20}  \\
Magnitude & $V$ (mag) & $9.52 \pm 0.03$ & \cite{SIMBAD} \\
System scale & $a/R_*$ & $4.08{}^{+0.02}_{-0.06}$ & \cite{Ehrenreich20} \\
Orbital period & $P$ (days) & $1.80988198{}^{+0.00000064}_{-0.00000056}$ & \cite{Ehrenreich20} \\
Transit duration & $T_{14}$ (min.) & $230$ & \cite{Ehrenreich20} \\
Epoch of mid-transit & $T_c$ (BJD) & $2458080.626165{}^{+0.000418}_{-0.000367}$ & \cite{Ehrenreich20} \\
Radius ratio & $R_\mathrm{p}/R_*$ & $0.10852 {}^{+0.00096}_{-0.00072}$ & \cite{Ehrenreich20}  \\
Planetary radius & $R_\mathrm{p}$ ($R_\mathrm{J}$) & $1.854{}^{+0.077}_{-0.076}$ & \cite{Tabernero20}  \\
Planetary mass & $M_\mathrm{p}$ ($M_\mathrm{J}$) & $0.894{}^{+0.014}_{-0.013}$ & \cite{Ehrenreich20} \\
Inclination & $i$ (degrees) & $89.623{}^{+0.005}_{-0.034}$ & \cite{Ehrenreich20} \\
Systemic velocity & $\gamma_\mathrm{sys}$ (km/s) & $-1.0733 \pm 0.0002$ & \cite{West16} \\
Stellar radial velocity semi-amplitude & K${}_*$  (m/s) & $116.02{}^{+1.29}_{-1.35}$ & \cite{Ehrenreich20} \\
Planetary radial velocity semi-amplitude & K${}_p$ (km/s) & $196.52 \pm 0.94$ & \cite{Ehrenreich20} \\
Quadratic limb darkening coefficient & $u_1$ & $0.393$ & \cite{Ehrenreich20} \\
Quadratic limb darkening coefficient & $u_2$ & $0.219$ & \cite{Ehrenreich20} \\
Projected equatorial rotational velocity & $v\sin i$ (km/s) & 1.48 $\pm$ 0.28 & \cite{Ehrenreich20} \\
\enddata
\end{deluxetable*}

\section{Data Reduction}
\label{sec:reduc}
The raw data frames were initially reduced with \texttt{OPERA}, the Open source Pipeline for ESPaDOnS Reduction and Analysis \citep{opera}. \texttt{OPERA} performs an optimal extraction, a bias subtraction, flat-fielding, a blaze correction, a continuum normalization, a wavelength calibration using arc lamps, and a correction to the wavelength solution using telluric lines. The result is a 1D calibrated spectrum for each exposure.

Next, we removed contaminating cosmic rays from the data through the use of a median absolute deviation filtering algorithm with a threshold of 5 times the median absolute deviation.
Following \cite{Allart17}, we then flux-scaled each spectrum by dividing out a reference spectrum (in this case chosen to be the first spectrum of the night) and fitting a fourth-order polynomial to a binned version of this ratio. We note that the choice of order does not significantly impact our results. This polynomial fit was then evaluated at the wavelength range of the unbinned spectra and divided out of each individual spectrum in order to re-normalize the data to the same continuum flux level. This process allows us to correct for flux level variations between individual exposures. The results of this data reduction process are presented in Appendix \ref{app:reduc}.

\subsection{Removal of Telluric and Stellar Features with \textsc{SysRem}}
\label{subsec:sysrem}
After the initial reduction, the spectra were dominated largely by telluric and stellar absorption features. We removed these using the \textsc{SysRem} algorithm \citep{Tamuz2005}, which has been well-established as a means of correcting telluric and stellar absorption that are essentially stationary in time \citep[e.g.,][]{Esteves17, Deibert19}. The large change in radial velocity of the exoplanet over the course of the transit ($\sim$ -52 km/s to $\sim$ +52 km/s in the case of WASP-76b) means that Doppler-shifted absorption features in the exoplanet's atmosphere are not affected.

To determine the optimum number of iterations of the \textsc{SysRem} algorithm to apply to our data, we interpolated our data to a common wavelength grid in the telluric rest frame (using the \texttt{scipy.interpolate} spline interpolation function; \citealt{2020SciPy-NMeth}). Note that this initial interpolation is necessary as \texttt{OPERA} corrects for potential drifts in the wavelength solution. We then ran between 1 and 10 iterations of the \textsc{SysRem} algorithm on relevant orders and carried out the data analysis routine described in Section \ref{subsec:transspec}. As in \cite{Turner20}, we chose to apply the number of iterations that maximized the significance of our detections, which in general was 6 iterations.
However, we note that the derived parameters for each detected species (see Section \ref{subsec:transspec}) were the same within the uncertainties regardless of the number of iterations applied.

\subsection{Correction for the Rossiter-McLaughlin Effect and Centre-to-Limb Variation}
\label{subsec:rm}
Finally, we corrected for the Rossiter-McLaughlin (RM) effect and  stellar center-to-limb variation (CLV) using the method outlined in \cite{Turner20}. We note that the star's $v\sin i$ value is low
\citep[1.48 $\pm$ 0.28 km/s,][]{Ehrenreich20},
so the magnitude of these effects (as calculated in \citealt{Tabernero20}) is not expected to be highly significant relative to the strength of our detections; nevertheless, we model and correct for them in order to ensure that they do not affect the reported signals.

To model the RM and CLV effects, angle-dependent spectra were generated using the \texttt{Spectroscopy Made Easy (SME)} package \citep[][]{SME} for 25 angles and the wavelength ranges matching those of the orders of interest for the stellar parameters taken from \cite{Ehrenreich20} (see Table~\ref{tab:parameters}). The star was modeled as a solid rotating body. The code uses a grid-based approach, with the stellar radius set to 510 pixels. The limb-angle dependent spectra are interpolated onto the stellar grid, taking the rotational velocity and limb-angle into account. The planet was modeled as a black disk. To ensure that the models were treated in a similar way to the data, they were continuum-normalized and subsequently divided by the out-of-transit model spectrum before being used to correct the data.

\section{Analysis}
\label{sec:analysis}

\subsection{Extraction of Transmission Spectra}
\label{subsec:transspec}
After reducing the data as described in Section \ref{sec:reduc}, we followed the methods of e.g., \cite{Wyttenbach15} and \cite{Turner20} to extract the planetary signal in the form of a transmission spectrum in the planet's rest frame. The adopted values for the systemic velocity $\gamma_{\text{sys}}$ and stellar radial velocity semi-amplitude K$_{*}$
used to shift to the stellar rest frame are presented in Table \ref{tab:parameters}, while the barycentric Earth radial velocity values were calculated with the \texttt{barycorr} applet \citep[][]{barycorr} using Julian Date values provided by \texttt{OPERA}.

For the Keplerian planetary radial velocity semi-amplitude K${}_p$, we tested two approaches: first, we used the value derived in \cite{Ehrenreich20} (see Table \ref{tab:parameters}); following this, we determined a value directly from our data for each detected line species (see Appendix \ref{app:pf}, with K${}_p$ values presented in Table \ref{tab:results}). We then determined an average K${}_p$ value for each species. In cases where the species was not detected, or where K${}_p$ could not be reliably extracted from the data, we only created the transmission spectrum with the \cite{Ehrenreich20} value.

Once the spectra were in the planet's rest frame, we summed over the in-transit frames to create the transmission spectrum. In general, we used all in-transit frames to create the transmission spectrum. However, we note that in the case of the sodium doublet, low SNR (signal-to-noise ratio) remnants of the stellar spectrum were present which may affect the extraction of the transmission spectrum (see Fig. \ref{fig:trail} in Appendix \ref{app:trail}). We thus follow e.g., \cite{Seidel20b} in fitting the stellar sodium lines (prior to the corrections described in Section \ref{subsec:sysrem}), calculating the full-width at half-maximum (FWHM) of each line, and masking out all points which fall within the FWHM of the lines before extracting the transmission spectrum. Across both lines, these masked values span 19 pixels.

\begin{figure*}
    \centering
    \includegraphics[]{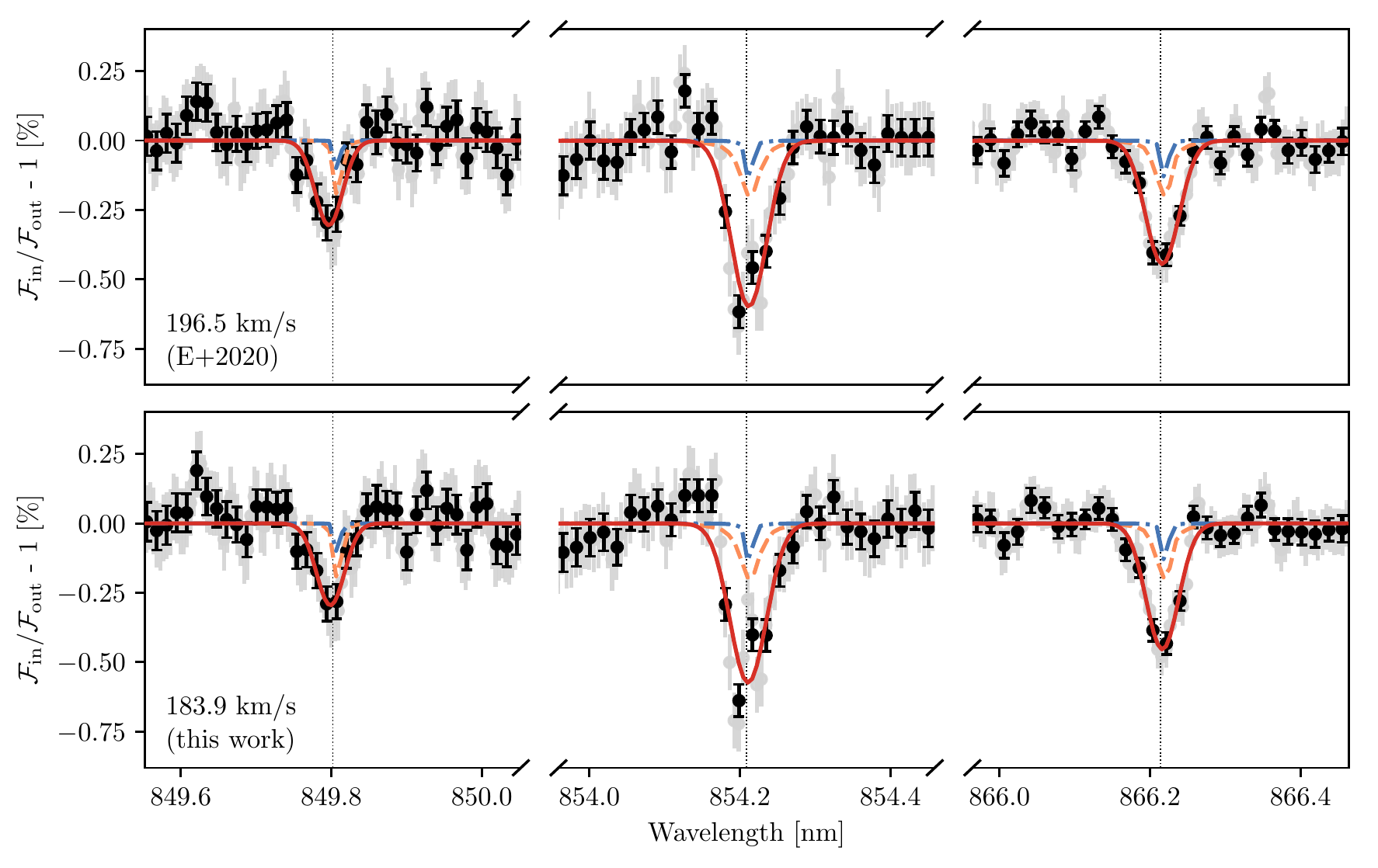}
    \caption{GRACES transmission spectra of WASP-76b around the three lines of the ionized calcium triplet in the planet rest frame. The top plot uses the K${}p$ value from \cite{Ehrenreich20} to shift to the planet rest frame, while the bottom plot uses the value extracted in this work. Note the breaks in the x-axes. The first line (at $\sim$ 849.8 nm) is located in the 27th order of the data, while the second and third lines (at $\sim$ 854.2 and 866.2 nm) are located in the 26th order. Grey points represent the transmission spectra (which have been normalized to the continuum), while black points represent the same spectra binned for clarity. The dotted black lines show the expected locations of each line in air wavelengths, and the solid red line shows a Gaussian fit to each individual line profile, with the fitted parameters of each Gaussian presented in Table \ref{tab:results}. The synthetic LTE and NLTE transmission spectra (see Section \ref{subsec:model}) are plotted in dot-dashed blue lines, and dashed orange lines, respectively.}
    \label{fig:calcium}
\end{figure*}

\begin{figure*}
    \centering
    \includegraphics[]{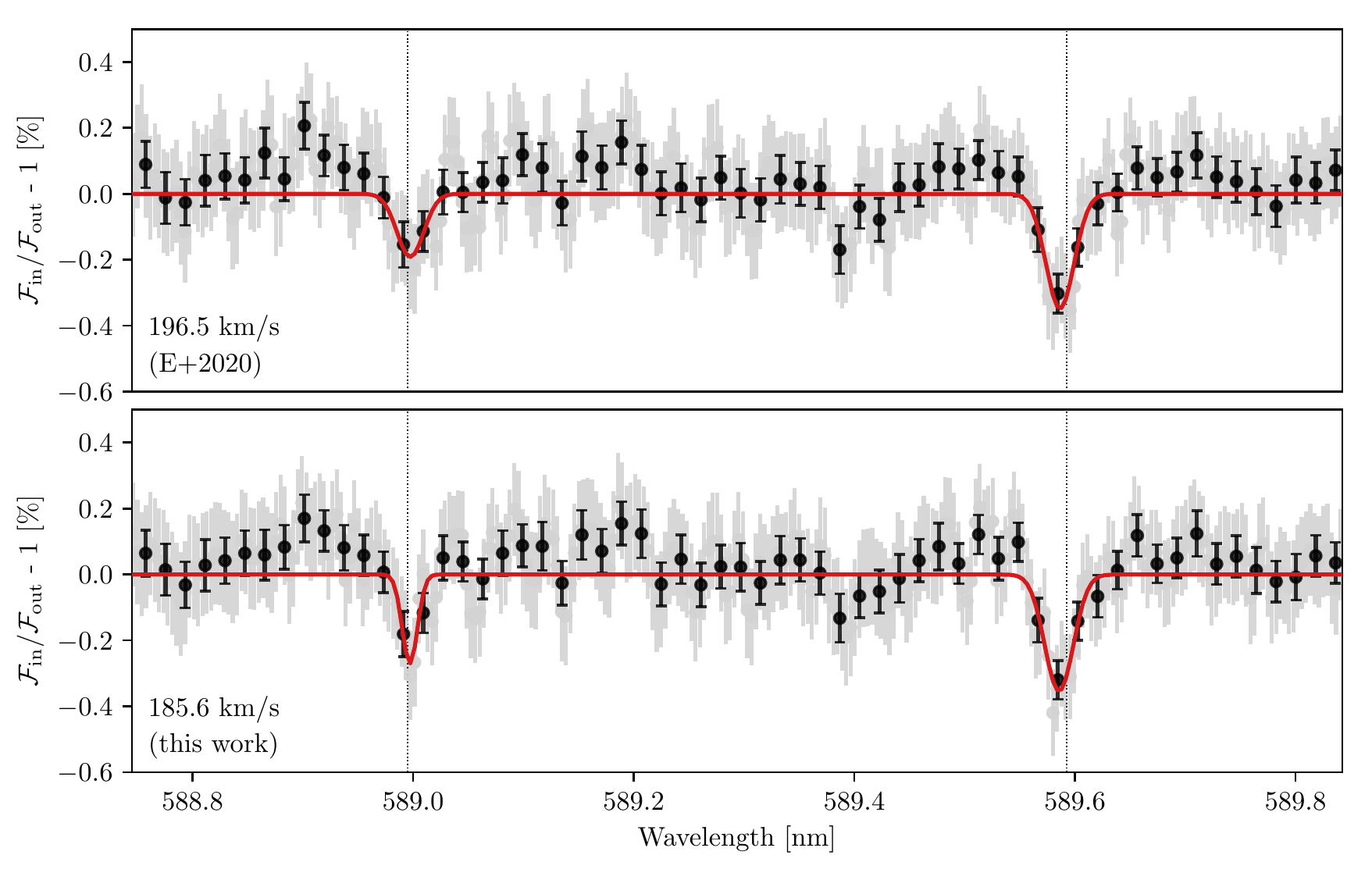}
    \caption{GRACES transmission spectra of WASP-76b around the sodium doublet, in the planet rest frame. As in Fig. \ref{fig:calcium}, the top plot uses the K${}p$ value from \cite{Ehrenreich20} to shift to the planet rest frame, while the bottom plot uses the value from this work. The two lines of the doublet are located near the edge of the 38th order. The data points and associated Gaussian fits are as described in the caption of Fig. \ref{fig:calcium}.}
    \label{fig:sodium}
\end{figure*}

The transmission spectra were generated separately for each relevant order, and are shown in Fig. \ref{fig:calcium} for the Ca~II triplet and Fig. \ref{fig:sodium} for the Na~I sodium doublet. Our tentative detections of Li~I and K~I and tentative non-detection of H$\alpha$ are presented in Appendix \ref{app:ha-li}.

As in \cite{Turner20}, we fit each atomic line with a Gaussian that describes the line center, absorption depth, and Gaussian standard deviation $\sigma$. We note that while most lines are fit individually, we fit the two lines of the Na~I doublet together. The fits were obtained with the \texttt{curve\_fit} function from the \texttt{scipy} library, with bounds applied to exclude non-physical values. From these parameters we determined an associated Doppler shift of the fitted line centre from the expected location of the line (which in some cases can be attributed to atmospheric winds) and the FWHM for each line (FWHM = $2\sqrt{2\ln2}\sigma$). Following \cite{Yan19}, we then used the fitted line depth $h$ and the photometric transit depth $\delta$ ($\sim$ 1.2\%; see Table \ref{tab:parameters}) to calculate the effective radius $R_\mathrm{eff}$ at each line center via:

\begin{equation}
\label{eq:reff}
    \frac{R_\mathrm{eff}^2}{R_p^2} = \frac{\delta + h}{\delta}
\end{equation}

The results are presented in Table \ref{tab:results}. We note that our results are generally consistent within the uncertainties regardless of whether we used our measured K${}_p$ values or the value derived by \cite{Ehrenreich20}. The exception is the depth of the Na~I D2 line (see Table \ref{tab:results}), which is slightly lower when the \cite{Ehrenreich20} value is used. This discrepancy could be due to noise in the data, as the D2 line is located near the edge of the spectral order, where the efficiency of the detector drops steeply and any variations in the blaze profile may impact the line shape. We also acknowledge that atmospheric dynamics may impact the recovery of K${}_p$ (e.g., see recent results and discussions from \citealt{Nugroho20} and \citealt{Wardenier21}), which is why we have confirmed that our results are consistent regardless of the value used.

\begin{deluxetable*}{ccccccc}
\label{tab:results}
\tablecaption{Summary of detected atmospheric line absorption parameters.}
\tablehead{%
    \colhead{Line} & \colhead{$\lambda$ [nm]} & \colhead{Depth [\%]} & \colhead{V${}_\mathrm{center}$ [km/s]} & \colhead{FWHM [km/s]} & \colhead{$R_\mathrm{eff}$ [$R_p$]} & \colhead{Extracted K${}_p$ [km/s]}
    }
\startdata
Na I D2 & 588.995 & 0.270 $\pm$ 0.062 & 1.043 $\pm$ 1.647 & 8.277 $\pm$ 2.180 & 1.11 $\pm$ 0.09 & -- \\
 & & (0.191 $\pm$ 0.048) & (1.258 $\pm$ 2.278) & (14.529 $\pm$ 4.198) & (1.08 $\pm$ 0.08) \\
Na I D1 & 589.592 & 0.354 $\pm$ 0.045 & -3.426 $\pm$ 0.963 & 15.520 $\pm$ 2.268 & 1.14 ${}^{+0.09}_{-0.08}$ & 185.6${}^{+12.2}_{-27.7}$ \\
 & & (0.348 $\pm$ 0.046) & (-3.015 $\pm$ 1.003) & (15.367 $\pm$ 2.362) & (1.14 ${}^{+0.09}_{-0.08}$) & \\
Ca II & 849.802 & 0.295 $\pm$ 0.034 & -1.081 $\pm$ 0.856 & 15.357 $\pm$ 2.016 & 1.12 $\pm$ 0.08 & -- \\
 & & (0.306 $\pm$ 0.033) & (-1.828 $\pm$ 0.815) & (15.156 $\pm$ 1.919) & (1.12 $\pm$ 0.08) & \\
Ca II & 854.209 & 0.574 $\pm$ 0.041 & 0.701 $\pm$ 0.697 & 20.000 $\pm$ 1.640 & 1.22 $\pm$ 0.09 & 184.3${}^{+11.8}_{-11.3}$ \\
 & & (0.599 $\pm$ 0.039) & (1.055 $\pm$ 0.625) & (19.542 $\pm$ 1.471) & (1.22 $\pm$ 0.09) & \\
Ca II & 866.214 & 0.454 $\pm$ 0.024 & 0.858 $\pm$ 0.445 & 17.532 $\pm$ 1.048 & 1.18 $\pm$ 0.08 & 183.5${}^{+8.9}_{-16.0}$ \\
 & & (0.445 $\pm$ 0.026) & (0.836 $\pm$ 0.518) & (18.098 $\pm$ 1.221) & (1.18 $\pm$ 0.08) & \\
\enddata
\tablecomments{For all lines, the first row represents the results obtained with the K${}_p$ value fit in this work (see Appendix \ref{app:pf}), while the second row (in brackets) represents the results obtained with the K${}_p$ value from \cite{Ehrenreich20}. Column 1: species detected. Column 2: expected location of species in air wavelengths. Column 3: fitted absorption depth of the line, assuming a Gaussian profile. Column 4: fitted offset from expected line location, assuming a Gaussian profile. In some cases, this could be attributed to atmospheric winds. Column 5: FWHM of the line (derived using a fitted value of $\sigma$) assuming a Gaussian profile. Column 6: effective radius at the line centre, calculated using Eq. \ref{eq:reff}. Column 7: Planetary radial velocity semi-amplitude extracted in this work. For lines without entries, we were unable to accurately extract a K${}_p$ value.}
\end{deluxetable*}

\subsection{Model of WASP-76b's Atmosphere}
\label{subsec:model}
To better assess the state of the atmosphere of WASP-76b, we turned to a range of modeling efforts. The atmospheric pressure-temperature (\emph{P--T}) profile was derived using the one-dimensional modeling methods described in \citet{Fortney08} and \citet{Fortney20}.  Planet-wide average conditions were assumed, as well as equilibrium chemistry at solar metallicity, and the model iterated to a solution in radiative-convective equilibrium.

Following the Gaussian fits described in the previous section, we further modeled the observed line profiles of the Ca~II triplet with synthetic Local Thermodynamic Equilibrium (LTE) and Non-Local Thermodynamic Equilibrium (NLTE) transmission spectra produced with the radiative transfer code \texttt{Cloudy} \citep[][]{Ferland17} on the basis of pressure-temperature and abundance profiles computed assuming radiative equilibrium. The scheme we employed is described by \citet{young2020} \citep[see also][]{Turner20,Fossati20,Fossati21}. However, to speed up the calculations, the synthetic transmission spectra have been computed accounting only for H, He, and Ca~II absorption. We note that the NLTE calculations are not self-consistent, since the pressure-temperature profile is derived assuming LTE and the pressure-temperature profile computed at the substellar point is assumed to be valid across the entire planet \citep[see][]{Fossati21}. Self-consistent NLTE calculations of the atmosphere of WASP-76b will be the subject of a future study.

\section{Results and Discussion}
\label{sec:discussion}
Table \ref{tab:results} presents the measured and derived line parameters for the atmospheric species detected through our analysis. Additional details on our tentative results are presented in Appendix \ref{app:ha-li}.

\subsection{Comparison with Previous Results}
\label{subsec:comparison}

The line depths we derived for the Na~I doublet (see Table \ref{tab:results}) are comparable within the uncertainties to those presented in \cite{Seidel19} (later updated in \citealt{Seidel21}) and \cite{Tabernero20}.

In the case of the Na~I D1 line, the measured offset of the line core from its expected location (i.e., the Doppler shift of the line, which could be attributed to atmospheric winds) is consistent with that reported by \cite{Tabernero20}. However, in the case of the D2 line center, we detect a redshift that is significantly different from the offset reported by \cite{Tabernero20} and that is unlikely to be physical; instead we attribute the shift to noise in the data. 
As mentioned previously, we expect this line to be noisier than the D1 line given its closer proximity to the edge of the spectral order. Note as well the large errors in these measurements.

In the case of the ionized calcium triplet, our results are comparable to those of \cite{Yan19}, who detected both the Ca~II H \& K lines and the triplet (with HARPS and CARMENES spectra, respectively) in the atmospheres of WASP-33b and KELT-9b. Their analysis revealed that the average line depths of the H \& K lines were significantly greater than those of the triplet in both cases. 
When comparing our detections to those of \cite{Tabernero20}, we find similar results: the Ca~II H and K lines were detected by \cite{Tabernero20} at line depths of 2.66 $\pm$ 0.29\% and 2.21 $\pm$ 0.20\% respectively (averaged between two transits).
In our case, the three lines of the Ca~II infrared triplet are detected at line depths of 0.295 $\pm$ 0.034\%, 0.574 $\pm$ 0.041\%, and 0.454 $\pm$ 0.024\% (using our K${}_p$ value; see Table \ref{tab:results}). The line depth ratios between the H \& K and triplet lines in the atmosphere of WASP-76b are thus consistent with those reported in WASP-33b and KELT-9b by \cite{Yan19}; this is to be expected, given that the H \& K lines are caused by resonant transitions from the ground state of Ca~II while the triplet is not. 

\subsection{Comparison with Synthetic LTE and NLTE Transmission Spectra}
\label{subsec:nlte}
Fig. \ref{fig:calcium} compares the detected Ca~II triplet with the synthetic LTE and NLTE transmission spectra described in Section \ref{subsec:model}. As can be seen in the figure, the observed line depths and derived FWHM (see Table \ref{tab:results}) are significantly larger than those predicted by the models, though the NLTE spectra provide a slightly closer fit in terms of both line widths and depths.

These results are similar to those of \cite{Yan19}, who used hydrostatic models (assuming isothermal temperatures and tidally-locked rotation) to simulate Ca~II absorption in the atmospheres of WASP-33b and KELT-9b and found that the observed line depths in both cases were significantly deeper than predicted. A recent detection of the Ca~II infrared triplet in the atmosphere of WASP-121b by \cite{Merritt21} was likewise best matched by an injected model that had been scaled to simulate significant upward motion in the planetary atmosphere. In our case, the large discrepancy between the predicted and observed line depths could also be attributed to the fact that the atmosphere extends further than the models assume, which could in turn be due to dynamics enhancing the density of calcium ions present at high altitudes.

Such a scenario is also consistent with the Ca~II H \& K line detections in the atmosphere of WASP-76b reported by \cite{Tabernero20}. The lines detected in that work have large intrinsic depths pointing towards high formation altitudes, and were suggested to originate from photo-ionization in the exoplanet's highly irradiated upper atmosphere. This same process could explain the large transit depths derived in our work.

We note as well that planetary winds and/or rotation could be broadening the detected lines; however, such a broadening may also be expected to decrease the line depths. While a full consideration of planetary wind patterns is beyond the scope of this work, a recent analysis by \cite{Seidel21} considered winds in WASP-76b's atmosphere in more detail.

Following \cite{Sing19}, we used the formalism of \cite{Gu03} to estimate an equivalent radius of the exoplanet's Roche lobe as viewed with transit geometry, under the assumption that the region probed by the Ca~II infrared triplet is in approximate hydrostatic equilibrium (though we note that this may not necessarily be the case). Using the parameters from Table \ref{tab:parameters}, we estimate a value of $1.41$ $R_p$. When this is compared to the effective radii calculated for our detected line profiles (Table \ref{tab:results}), we see that the atomic species probed by our observations are consistent (within the errors) of being close to, but still within, the exoplanet's Roche radius.
As in \cite{Yan19} and \cite{Turner20}, we thus infer that we are detecting ionized calcium from the upper extended atmosphere (i.e., within the Roche radius) rather than material which has already escaped.

While the presence of H$\alpha$ absorption might shed further light on hydrodynamic outflows in the planet's upper atmosphere, previous searches for H$\alpha$ have proved inconclusive and pointed to potential atmospheric variability. \cite{Tabernero20} detected H$\alpha$ in one transit, but their second transit did not yield a clear detection. Using HST spectra, \cite{vonEssen20} were also unable to detect strong H$\alpha$ absorption. Together, these results suggest that atomic hydrogen may be variable in the atmosphere of WASP-76b.
Our own search for H$\alpha$ was similarly inconclusive (see Appendix \ref{app:ha-li}); this could be evidence for atmospheric variability, but additional data are needed to confirm. 

Although it is likely that WASP-76b indeed hosts an escaping atmosphere and that hydrodynamics affect the line shapes, it is also likely that the layers of the planetary atmosphere probed by the Ca~II lines are significantly hotter than predicted by radiative equilibrium. The higher temperature would then lead to enhanced production of ionized calcium and thus to strong absorption features \citep[e.g.,][]{Turner20}. Identified mechanisms leading to increases in temperature in the atmospheres of ultra-hot Jupiters are metal photoionisation \citep[][]{Lothringer18,Fossati20} and NLTE effects in the form of overpopulation of species responsible for heating (e.g., Fe~II) together with underpopulation of species responsible for cooling \citep[e.g., Fe~I and Mg,][]{Fossati21}. Indeed, \cite{Seidel21} retrieved temperature upper limits significantly higher than T$_{\text{eq}}$ from the sodium doublet, further supporting this possibility.

In a follow-up study, we will explore the capabilities of self-consistent modeling, accounting for metal photoionisation and NLTE effects in the computation of both the \emph{P--T} profile and transmission spectrum, to match the observations. Such an investigation will provide a stronger basis on which to consider the effect of hydrodynamics in the middle and upper atmosphere of WASP-76b.

\section{Conclusion}
\label{sec:conc}
We analyzed transit observations of WASP-76b taken with GRACES at the Gemini North Telescope as part of the Large and Long Program GN-2020B-LP-106: ``Exploring the Diversity of Exoplanet Atmospheres at High Spectral Resolution'' (ExoGemS),
resulting in detections of Na~I and Ca~II. The latter represents the first detection of the ionized calcium triplet at $\sim$ 850 nm in the atmosphere of WASP-76b, and the fifth such detection in ultra-hot Jupiters. 

In a forthcoming paper, we will analyze the full spectral range covered by these observations and search for absorption due to a suite of atomic, ionic and molecular species modelled at high resolution (following \citealt{Hood20}), including the species recently detected by \cite{Ehrenreich20} and \cite{Tabernero20}.

\acknowledgements

We thank Miranda Herman and Tom\'{a}s Cassanelli for the helpful discussions. We also thank Matteo Brogi, Drake Deming, Miranda Herman, and David Lafreni\`ere for their contributions as Co-Investigators of the ExoGemS survey. Finally, we thank the scientific editor Fred Rasio and the anonymous referee for their swift and constructive feedback on this work.

EKD is supported by a Vanier Canada Graduate Scholarship - NSERC.
JJF and CEH acknowledge the support of NASA Exoplanets Research Program grant 80NSSC19K0293. RA is a Trottier Postdoctoral Fellow and acknowledges support from the Trottier Family Foundation. This work was supported in part through a grant from FRQNT.

This work was based on observations obtained through the Gemini Remote Access to CFHT ESPaDOnS Spectrograph (GRACES). ESPaDOnS is located at the Canada-France-Hawaii Telescope (CFHT), which is operated by the National Research Council of Canada, the Institut National des Sciences de l'Univers of the Centre National de la Recherche Scientifique of France, and the University of Hawai'i. ESPaDOnS is a collaborative project funded by France (CNRS, MENESR, OMP, LATT), Canada (NSERC), CFHT and ESA. ESPaDOnS was remotely controlled from the international Gemini Observatory, a program of NSF's NOIRLab, which is managed by the Association of Universities for Research in Astronomy (AURA) under a cooperative agreement with the National Science Foundation on behalf of the Gemini partnership: the National Science Foundation (United States), National Research Council (Canada), Agencia Nacional de Investigaci\'{o}n y Desarrollo (Chile), Ministerio de Ciencia, Tecnolog\'{i}a e Innovaci\'{o}n (Argentina), Minist\'{e}rio da Ci\^{e}ncia, Tecnologia, Inova\c{c}\~{o}es e Comunica\c{c}\~{o}es (Brazil), and Korea Astronomy and Space Science Institute (Republic of Korea).

This work was enabled by observations made from the Gemini North telescope, located within the Maunakea Science Reserve and adjacent to the summit of Maunakea. We are grateful for the privilege of observing the Universe from a place that is unique in both its astronomical quality and its cultural significance.

This research also made use of the NIST Atomic Spectra Database funded [in part] by NIST’s Standard Reference Data Program (SRDP) and by NIST’s Systems Integration for Manufacturing Applications (SIMA) Program.

A subset of the figures in this work were generated with colorblind safe color schemes provided by ColorBrewer 2.0 \citep[][]{ColorBrewer}. The original ColorBrewer (v1.0) was funded by the NSF Digital Government program during 2001-02, and was designed at the GeoVISTA Center at Penn State (National Science Foundation Grant No. 9983451, 9983459, 9983461). The design and rebuilding of this new version (v2.0) was donated by Axis Maps LLC, winter 2009 and updated in 2013.

\facilities{Gemini:Gillett (GRACES), CFHT (ESPaDOnS)}

\software{
astropy \citep{astropy:2018},
barycorr \citep{barycorr}
batman \citep{batman},
Cloudy \citep{Ferland17},
ColorBrewer 2.0 \citep{ColorBrewer},
IPython \citep{4160251},
Matplotlib \citep{Hunter:2007},
Numpy \citep{harris2020},
OPERA \citep{opera},
SciPy \citep{2020SciPy-NMeth},
SME \citep{SME}
}

\vspace{\baselineskip}

\begin{appendix}
\section{Observing Conditions}
\label{app:obscond}
We observed one transit of WASP-76b with GRACES at the Gemini North telescope. The spectra cover the optical wavelengths from 400 to 1050 nm, and the resolution of the observations ranged from R $\sim$ 58,000 to $\sim$ 64,000. 

In total, we collected 169 spectra which included 120 spectra in-transit and 49 spectra out-of-transit. The observations spanned a period of $\sim$ 5.16 hours. We note that approximately 10\% of the transit, covering orbital phases of roughly 0.005 to 0.012, was lost due to technical issues at the observatory. 

\begin{figure}
    \centering
    \includegraphics{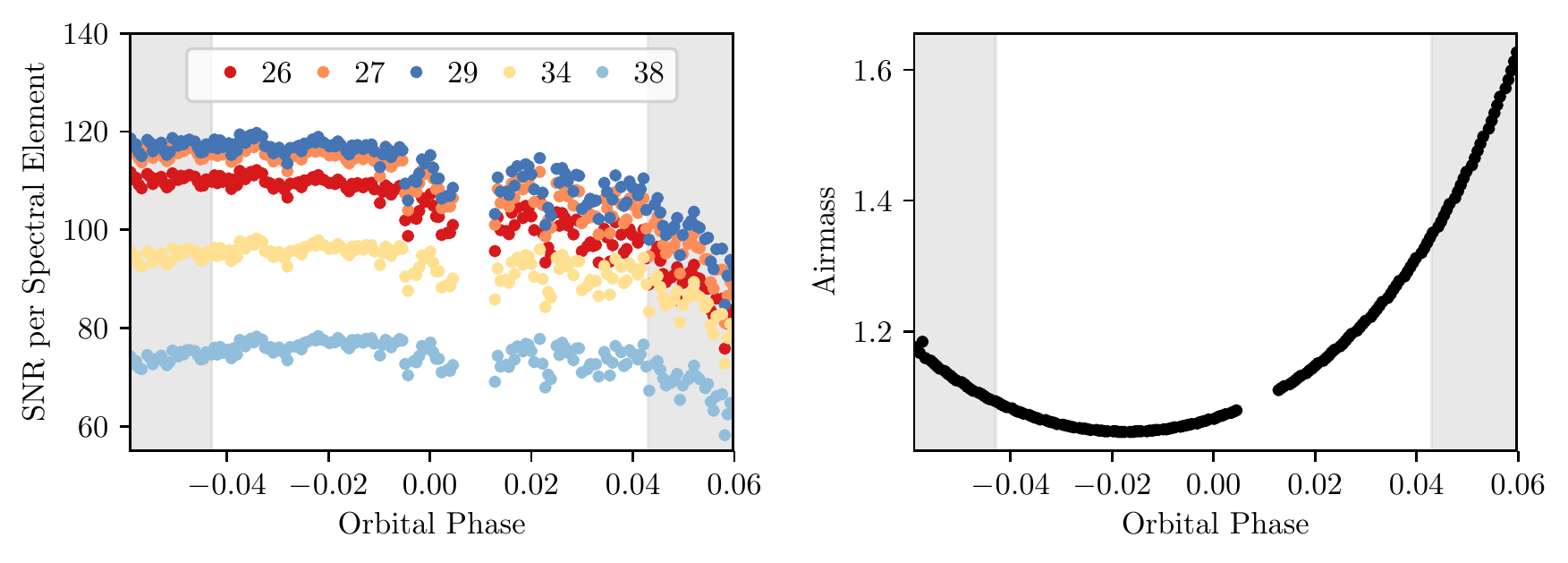}
    \caption{A summary of the observing conditions throughout the night. In both plots, out-of-transit phases are indicated by shaded grey regions. Left: the SNR per spectral element (at the centre of the order) for each order considered in this analysis. Order 26 contains two lines of the Ca~II triplet, order 27 contains one line of the Ca~II triplet, order 29 contains K~I, order 34 contains H$\alpha$ and Li~I, and order 38 contains the Na~I doublet. Right: the variation in airmass throughout the observing period. In both plots, note the gap in the data corresponding to observations that were lost due to technical issues at the observatory.}
    \label{fig:obs}
\end{figure}

Each spectrum was recorded with an exposure time of 60 seconds, resulting in average signal-to-noise ratios (SNR) per spectral element varying between $\sim$ 73.8 and $\sim$ 111.1 for the orders relevant to this work. The change in SNR throughout the night for each relevant order is shown in the left panel of Fig. \ref{fig:obs}. The airmass varied between 1.046 and 1.627 over the course of the observations (see the right panel of Fig. \ref{fig:obs}). 

\section{Data Reduction}
\label{app:reduc}

\begin{figure}
    \centering
    \includegraphics{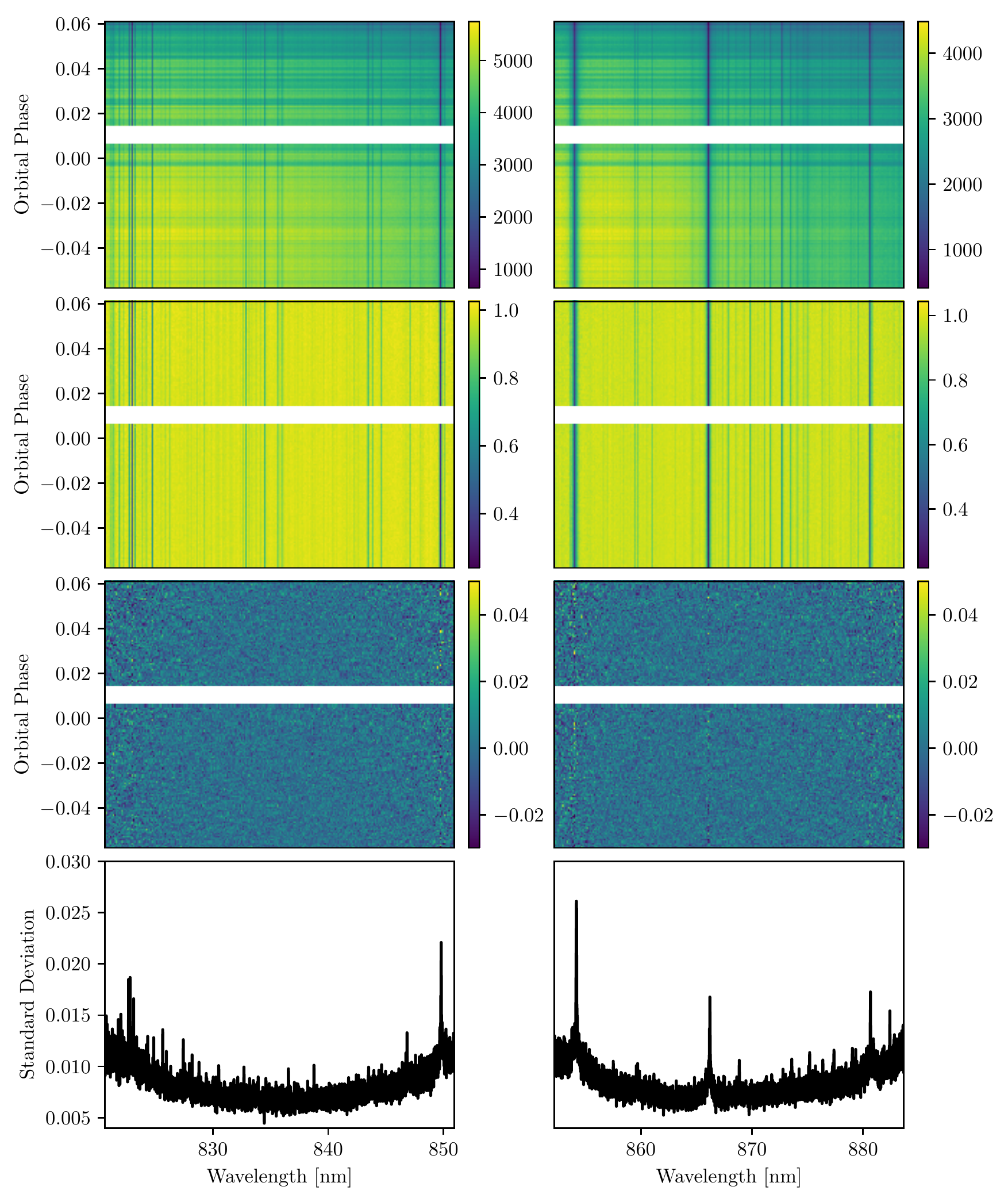}
    \caption{The full data reduction routine described in Section \ref{sec:reduc}, applied to the two orders containing the Ca~II triplet: order 27 (left) and order 26 (right). Note the gaps in the observations corresponding to technical difficulties at the observatory. For both columns, the top row shows the raw, unnormalized spectra after extraction from the telescope (note the scale of the colorbar, which represents unnormalized flux values). The second row presents the results of applying the flux normalization and cosmic ray removal routines. The third row shows the results of applying 6 iterations of the \textsc{SysRem} algorithm to our data (see Section \ref{subsec:sysrem}), and the bottom row shows the standard deviation along each wavelength channel after applying \textsc{SysRem}.}
    \label{fig:reduc-calcium}
\end{figure}

\begin{figure}
    \centering
    \includegraphics{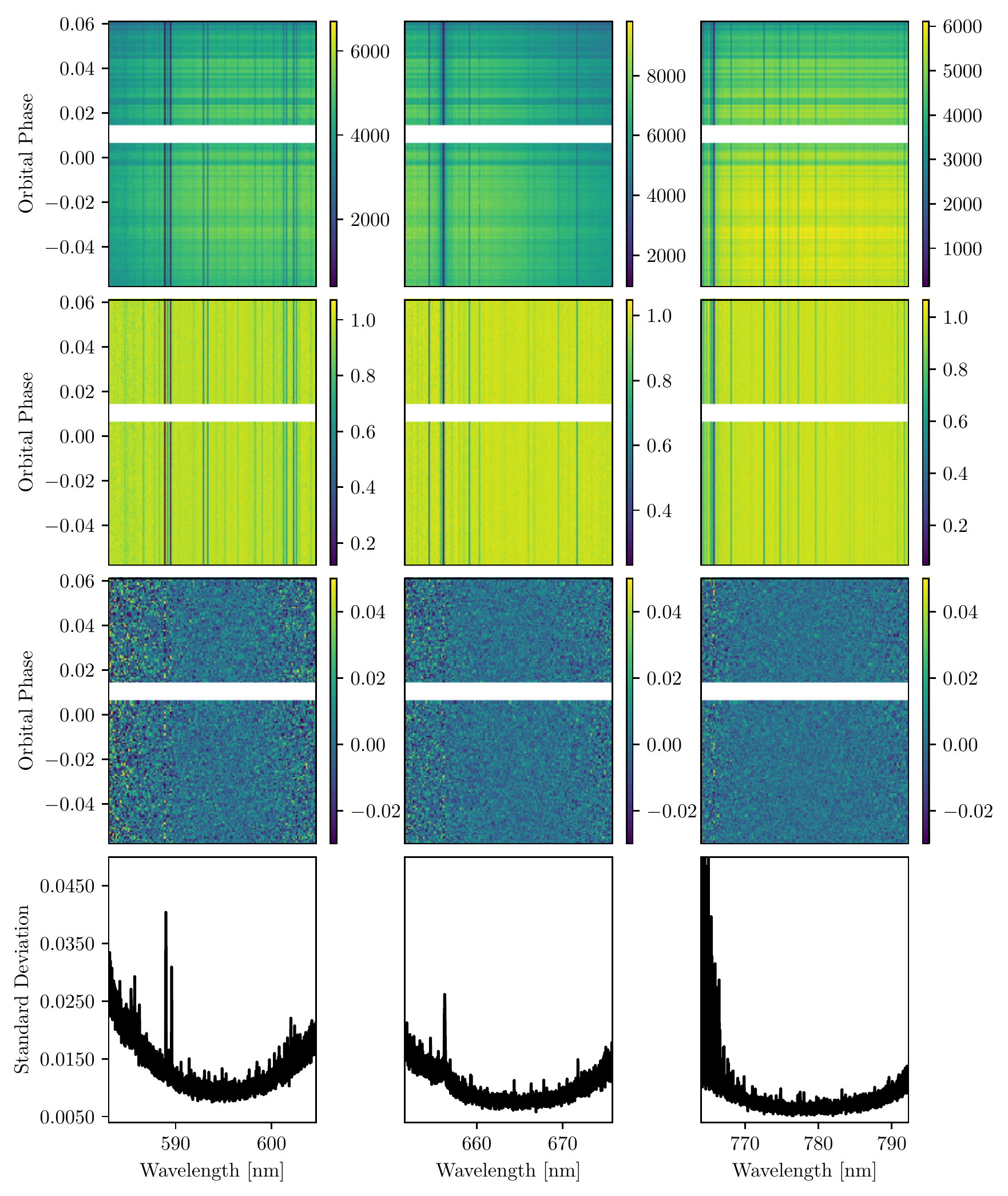}
    \caption{The full data reduction routine applied to the orders containing the Na~I doublet (order 38, left), H$\alpha$ and Li~I (order 34, middle), and K~I (order 29, right). The rows are as described in the caption of Fig. \ref{fig:reduc-calcium}.}
    \label{fig:reduc-other}
\end{figure}

The individual steps of the data reduction process described in Section \ref{sec:reduc}, including the correction for telluric and stellar absorption using the \textsc{SysRem} algorithm (Section \ref{subsec:sysrem}), are presented in Fig. \ref{fig:reduc-calcium} for the two orders containing the Ca~II infrared triplet and Fig. \ref{fig:reduc-other} for the three other orders considered in this analysis.

\section{Extracting Planetary Radial Velocity Semi-Amplitudes}
\label{app:extract}
\subsection{2D Phase-Folded Maps}
\label{app:pf}
\begin{figure}
    \centering
    \includegraphics{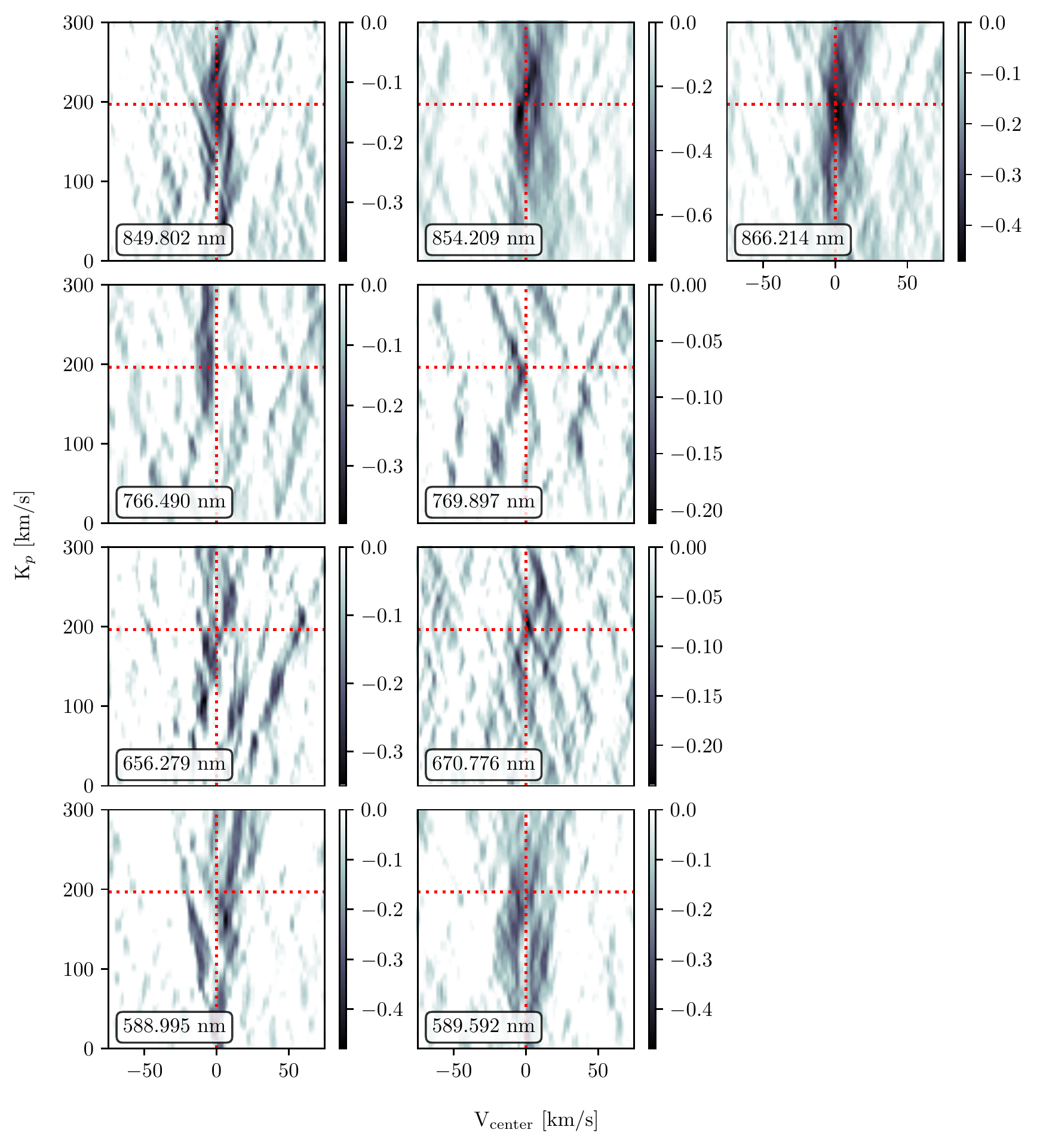}
    \caption{Phase-folded transmission spectra for all lines considered in this work. Each spectrum was phase-folded to radial velocities ranging from 0 km/s to 300 km/s in steps of 0.1 km/s. The air rest wavelength of each line is displayed in the bottom left of each plot, and the dotted red lines indicate the expected K${}_p$ (based on the value derived by \cite{Ehrenreich20}; see Table \ref{tab:parameters}) and V$_{\text{center}}$, i.e., 0 km/s. The colorbars indicate absorption depth in percentages. The top row shows the three lines of the Ca~II infrared triplet; the second row shows the two lines of the K~I doublet; the third row shows H$\alpha$ and Li~I (left and right respectively), and the bottom row shows the two lines of the Na~I doublet.}
    \label{fig:kp-vsys}
\end{figure}

Fig. \ref{fig:kp-vsys} shows the results of phase-folding the transmission spectra in order to determine the radial velocity semi-amplitude for each line. After shifting the data to the stellar rest frame, each spectrum was phase-folded to potential planetary radial velocity semi-amplitudes ranging from 0 km/s to 300 km/s, with a step size of 0.1 km/s, to create the K${}_p$-V$_{\text{center}}$ plots. Note that while the planetary radial velocity semi-amplitude is known to a far greater precision than this from previous studies, we sampled a wide range of radial velocities in order to investigate any potential peaks due to noise at other locations in the data (e.g., a correlation with residual stellar absorption might be expected at a K${}_p$ value of 0 km/s in the K${}_p$-V$_{\text{center}}$ map).
The peaks of the phase-folded K${}_p$-V$_{\text{center}}$ plots were then used to determine the planetary radial velocity semi-amplitudes, which are presented in Table \ref{tab:results}. Errors were estimated by determining the point by which this peak velocity value had dropped off by 1-$\sigma$; these errors are in general large enough that individual lines in a species are consistent with one another, as well as with values reported in previous works (e.g., K$_p$ = $196.52 \pm 0.94$ \citep{Ehrenreich20}, though we note that our values are in general lower).
In each plot, a V$_{\text{center}}$ value of 0 km/s corresponds to the expected location of each line, because the data have already been shifted to the stellar rest frame.

We note that there were several additional peaks of comparable strengths to the planetary signal in the K${}_p$-V$_{\text{center}}$ map created for the first line of the Ca~II infrared triplet (i.e., the line at $\sim$ 849.8; see the top left panel of Fig. \ref{fig:kp-vsys}). However, given the proximity of this line to the edge of the spectral order, as well as the fact that this line is the weakest of the three in the triplet, we assume that this is caused by noise present in the data and adopt the peak K${}_p$ value consistent with values determined for the other two lines (see Table \ref{tab:results}). We also note that the the peak K${}_p$ value of the Na~I D2 line is significantly lower than expected (though there is also an absorption feature present at the expected location); we likewise attribute this to noise at the edge of the spectral order and adopt the measured peak value of the Na~I D1 line for the sodium transmission spectrum. In the case of H$\alpha$, Li~I, and K~I, we are unable to recover a peak value from the K${}_p$-V$_{\text{center}}$ maps. We thus only use the \cite{Ehrenreich20} value to create these transmission spectra.

Finally, we note that the broad signals present in these plots are comparable to that seen in Fig. B.1 of \cite{Seidel21}.

\subsection{Spectra in Planetary Rest Frame}
\label{app:trail}

\begin{figure}
    \centering
    \includegraphics{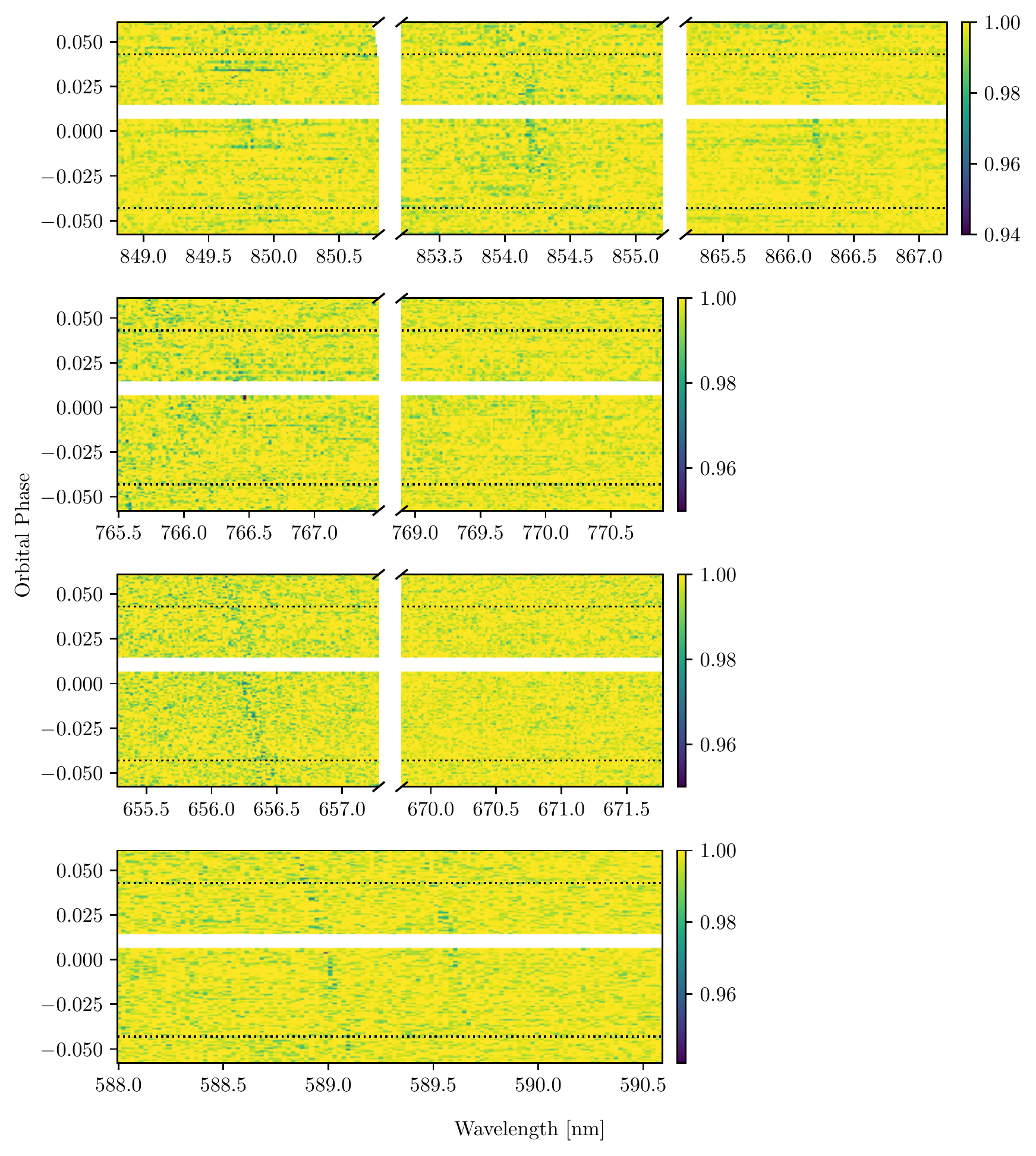}
    \caption{The WASP-76b spectra in the planetary rest frame, after the data reduction steps outlined in Section \ref{sec:reduc}. The colorbars correspond to normalized flux values. The top row shows the three lines of the Ca~II infrared triplet; the second row shows the two lines of the K~I doublet; the third row shows H$\alpha$ and Li~I (left and right respectively); and the final row shows the Na~I doublet. Note the various gaps in the x-axes, and the missing data corresponding to technical difficulties at the observatory. The dotted black lines represent points of first and fourth contact. In all cases, the data have been binned 4x for visualization purposes. At this point, any planetary absorption would be present in the form of a vertical line at the expected line locations. 
    Such lines can be seen by eye in the top middle and top right panels for the Ca~II lin at $\sim$ 854.2 nm and $\sim$ 866.2 nm respectively. In the case of H$\alpha$, some residual stellar absorption is visible; this could be contributing to the non-detection presented in Appendix \ref{app:ha-li}.}
    \label{fig:trail}
\end{figure}

Fig. \ref{fig:trail} shows the spectra in the planetary rest frame, after the data reduction steps outlined in Section \ref{sec:reduc} and using the K${}_p$ values measured in Appendix \ref{app:pf} (or in cases where K${}_p$ could not be reliably measured, using the value from \citealt{Ehrenreich20}; see Table \ref{tab:parameters}). In theory, a strong planetary signal could be visible by eye in the form of a vertical line at or near the expected absorption line location. In practice, while such lines are marginally visible in the case of the two strongest Ca~II lines, summing over in-transit frames and creating transmission spectra (as described in Section \ref{sec:analysis}) increases the signal strength to the point that the planetary absorption can be seen (e.g., Figs. \ref{fig:calcium} and \ref{fig:sodium}). 

\section{Hydrogen, Lithium, and Potassium Transmission Spectra}
\label{app:ha-li}
\begin{figure}
    \centering
    \includegraphics{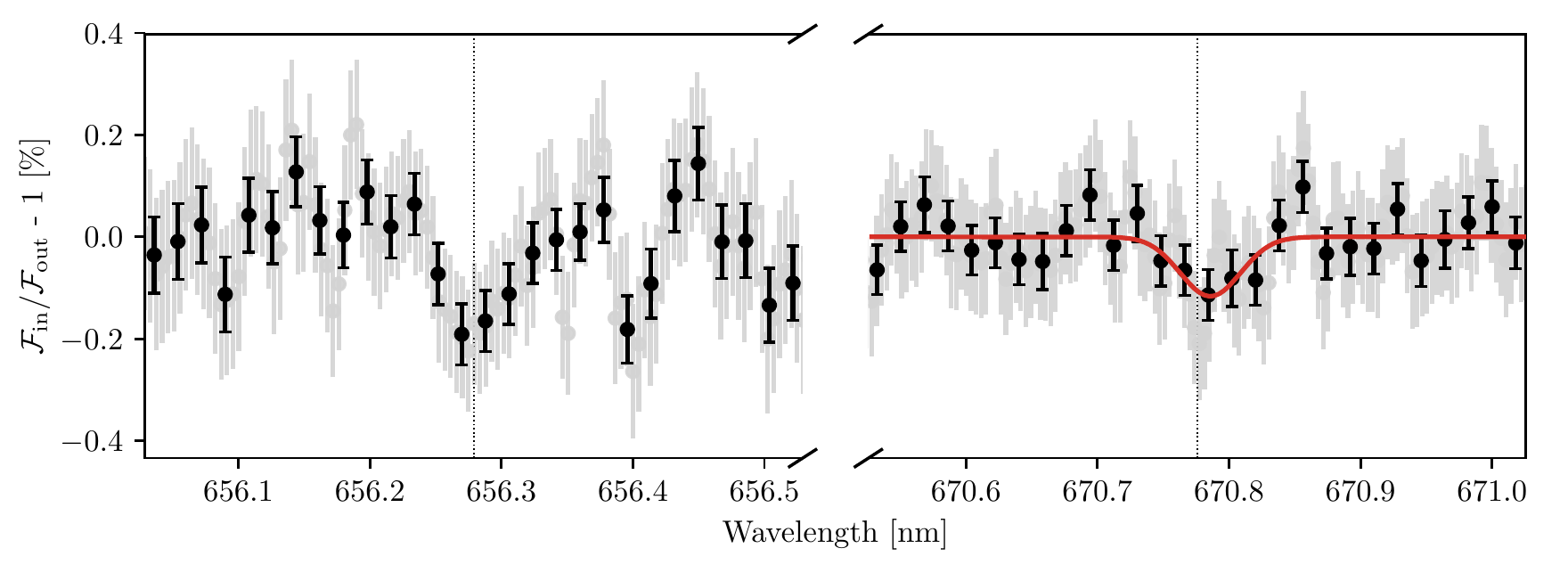}
    \caption{GRACES transmission spectra of WASP-76b around the expected locations of H$\alpha$ (left) and Li~I (right) in the planet rest frame. Note the breaks in the x-axis. Both lines are located in the 34th order of the data. Grey points represent the transmission spectra (which have been normalized to the continuum), while black points represent the same spectra binned for clarity. The dotted black lines show the expected locations of each absorption line in air wavelengths, and the solid red lines show Gaussian fits to the line profiles. While both spectra do show dips at the expected line locations, the noise levels in the spectra are too high to confidently classify these as detections.}
    \label{fig:ha-li}
\end{figure}

\begin{figure}
    \centering
    \includegraphics{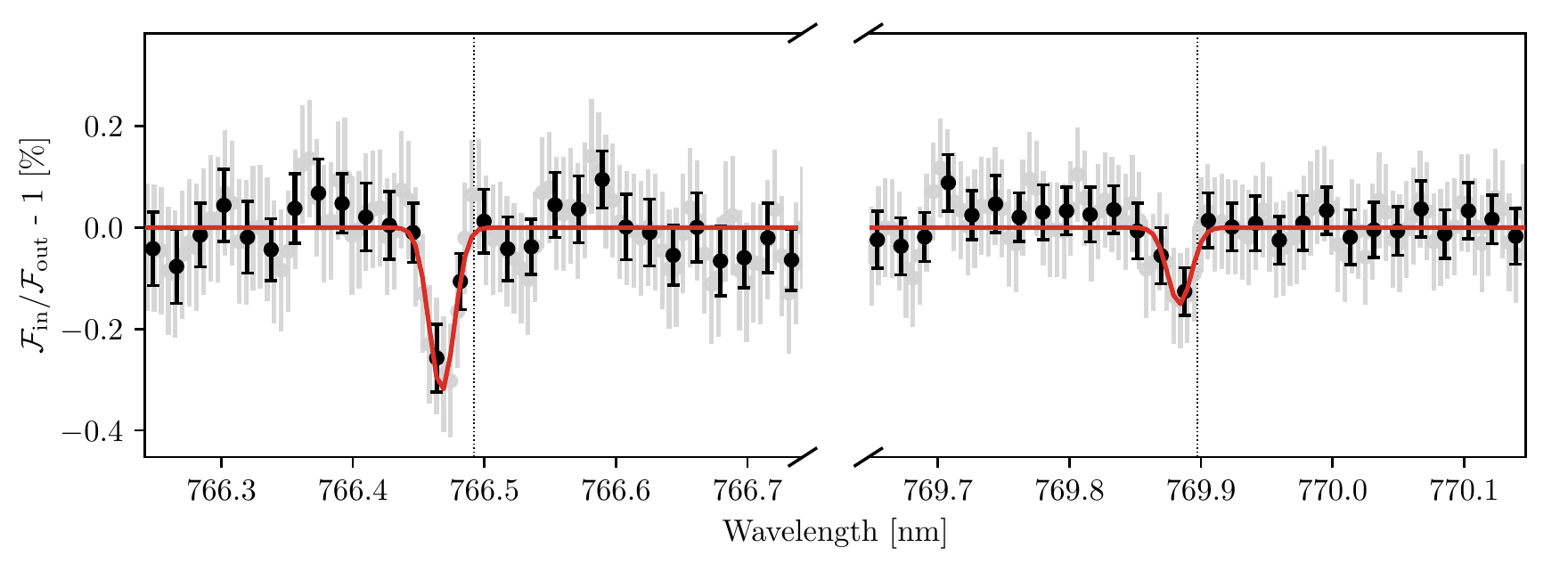}
    \caption{GRACES transmission spectra of WASP-76b around the expected locations of the two lines of the K~I doublet, in the planet rest frame. The plots are as described in the caption of Fig. \ref{fig:ha-li}.}
    \label{fig:k}
\end{figure}

The transmission spectra generated for the order containing H$\alpha$ and Li~I are presented in Fig. \ref{fig:ha-li}, while the transmission spectra generated for the two lines of the K~I doublet are presented in Fig. \ref{fig:k}. The process used to generate the spectra is the same as that outlined in Section \ref{subsec:transspec}.

In the case of H$\alpha$, we note that while there is absorption present in the spectrum at the expected line location, its depth is comparable to the overall noise levels in the spectrum (e.g., note the dip in the spectrum at $\sim$ 656.4 nm) and thus could be due to noise rather than any potential planetary signal. This result is largely consistent with those presented in \cite{Tabernero20}. In particular, the upper limit inferred for the second H$\alpha$ transit analyzed by \cite{Tabernero20} is at the noise level of our H$\alpha$ transmission spectrum, while the absorption depth measured for their first transit (0.48\% $\pm$ 0.12) is at/above the noise level of our spectrum. This could indicate (as was inferred in \citealt{Tabernero20}) that H$\alpha$ may be variable in the atmosphere of WASP-76b, but further data are needed to confirm this scenario.

We note as well that a low SNR stellar H$\alpha$ remnant is present in the data (see Fig. \ref{fig:trail} in Appendix \ref{app:trail}). The broad nature of the stellar H$\alpha$ line means that this low SNR remnant overlaps with the expected location of planetary absorption across essentially all in-transit frames; we thus do not mask these pixels as we did for the sodium doublet (see Section \ref{subsec:transspec}), but note that the contamination from this low SNR remnant is likely affecting our ability to confidently extract a transmission spectrum. 

In the cases of Li~I and K~I, the noise levels are slightly lower than the dips present at the expected line locations (note that Li~I is in the same spectral order as H$\alpha$ but is located closer to the center of the order; see Fig. \ref{fig:reduc-other}). We are able to fit Gaussians to each line profile and derive associated parameters as described in Section \ref{subsec:transspec}; these are presented in Table \ref{tab:results-li-k}. However, we note that in all cases there are various dips/peaks in the spectra that approach the depths of the fitted line profiles, which could call the validity of these results into question. We thus label these as tentative detections warranting further investigation, and provide additional insights through the bootstrapping analysis described in Appendix \ref{app:emc}. 

\begin{deluxetable}{ccccccc}
\label{tab:results-li-k}
\tablecaption{Summary of tentative atmospheric line absorption parameters. Note that unlike in Table \ref{tab:results}, we do not present measured K${}_p$ values; the line depths in this case were not strong enough to be extracted from the K${}_p$-V$_{\text{center}}$ maps. We thus only create tranmission spectra using the K${}_p$ value from \cite{Ehrenreich20}.}
\tablehead{%
    \colhead{Line} & \colhead{$\lambda$ [nm]} & \colhead{Depth [\%]} & \colhead{V$_{\mathrm{center}}$ [km/s]} & \colhead{FWHM [km/s]} & \colhead{$R_{\mathrm{eff}}$ [$R_p$]}
    }
\startdata
Li I & 670.776 & 0.117 $\pm$ 0.024 & 4.484 $\pm$ 2.435 & 24.160 $\pm$ 5.734 & 1.05 $\pm$ 0.08 \\
K I & 766.492 & 0.321 $\pm$ 0.042 & $-9.552$ $\pm$ 0.559 & 8.748 $\pm$ 1.317 & 1.13 $\pm$ 0.08 \\
K I & 769.897 & 0.151 $\pm$ 0.028 & $-5.254$ $\pm$ 0.754 & 8.285 $\pm$ 1.773 & 1.06 $\pm$ 0.08 \\
\enddata
\tablecomments{Column 1: species detected. Column 2: expected location of species in air wavelengths. Column 3: fitted absorption depth of the line, assuming a Gaussian profile. Column 4: fitted offset from expected line location, assuming a Gaussian profile. In some cases, this could be attributed to atmospheric winds. Column 5: FWHM of the line (derived using a fitted value of $\sigma$) assuming a Gaussian profile. Column 6: effective radius at the line centre, calculated using Eq. \ref{eq:reff}.}
\end{deluxetable}

Despite only being tentative, the measured line profiles for Li~I and K~I presented in Figs. \ref{fig:ha-li} and \ref{fig:k} are largely consistent with those in \cite{Tabernero20}. In particular, our detected line depths are comparable to those of \cite{Tabernero20}. Our measured blueshift for the K~I line at $\sim$ 769.9 nm (which could be consistent with atmospheric winds) is significantly different from that reported in \cite{Tabernero20}; however, we note that the values reported by \cite{Tabernero20} for each of their transits are also significantly different from each other ($-10.1$ $\pm$ 2.7 km/s and $-2.2$ $\pm$ 2.1 km/s). Our measured blueshift of the K~I line at $\sim$ 766.5 nm is consistent with that of the first transit analyzed in \cite{Tabernero20}; however, we note that this line falls within a dense forest of telluric O${}_2$ lines and thus may have been affected by the telluric absorption correction (in particular, a pair of strong telluric O${}_2$ lines are located directly bluewards of the K~I line, perhaps contributing to the absorption depth). We note as well that \cite{Tabernero20} did not investigate the presence of this line. While the K~I line at $\sim$ 766.5 nm does appear to be a somewhat clearer detection, the fact that we are unable to confidently recover the line at $\sim$ 769.9 nm indicates that systematic effects may be impacting our results, and we thus classify K~I as a tentative detection warranting further investigation. The large difference in velocity shifts between the two lines also calls the validity of these tentative results into question, as they would be expected to be consistent with one another.
We note that the telluric oxygen lines in this region could significantly affect the line profile of the lines, resulting in spurious shifts.

Finally, the large redshift that we measure for the Li~I is unlikely to be physical, and could further suggest that the result is indeed spurious, or at the least, affected by noise in the data.

\section{Systematic Effects and Robustness of Results: Empirical Monte Carlo Analysis}
\label{app:emc}

To further assess the contribution of systematic errors to our results, we carried out a bootstrapping or Empirical Monte Carlo (EMC) analysis \citep[e.g.,][among others]{Redfield08, Seidel19, CB19, Allart20}. For each line, we created three different scenarios: an out-out scenario where we randomly divided all out-of-transit frames into ``virtual'' in-transit and out-of-transit frames and repeated our analysis; an in-in scenario where we did the same but with in-transit frames; and an in-out scenario, where an increasing number of in-transit frames (up to half of the sample) were randomly removed from the sample and used to create a virtual in-transit data set that was compared with the nominal out-of-transit sample. For the first two scenarios, the distribution of derived absorption depths is expected to be centered at zero, while for the latter we expect to see a distribution centered at negative values for detected species and a distribution centered at 0 for non-detections. Note that the widths of the distributions may vary due to the different number of spectra used in each case; this does not affect the analysis. In all cases, we ran 10,000 trials of each scenario. 

Absorption depths were calculated via the method described in \cite{Redfield08}, \cite{Seidel19}, and \cite{Allart20}, among others. The average flux in a narrow (0.75 \AA{}) window around the line of interest was compared to the average flux in blue and red passbands outside the area of interest. This is different from the process used to derive absorption line parameters in our main analysis (i.e., fitting a Gaussian), but allowed us to more readily account for cases where absorption depths of 0\% were detected but the spectra themselves were noisy (note that a Gaussian fitting algorithm may attempt to fit on a noise spike, and checking each of the 10,000 trials by eye for proper fits was not feasible). We note that in the cases of the Na~I D2 and K~I lines, we centered the narrow flux windows around the detected locations of the lines accounting for Doppler shifts. 

\begin{figure}
    \centering
    \includegraphics{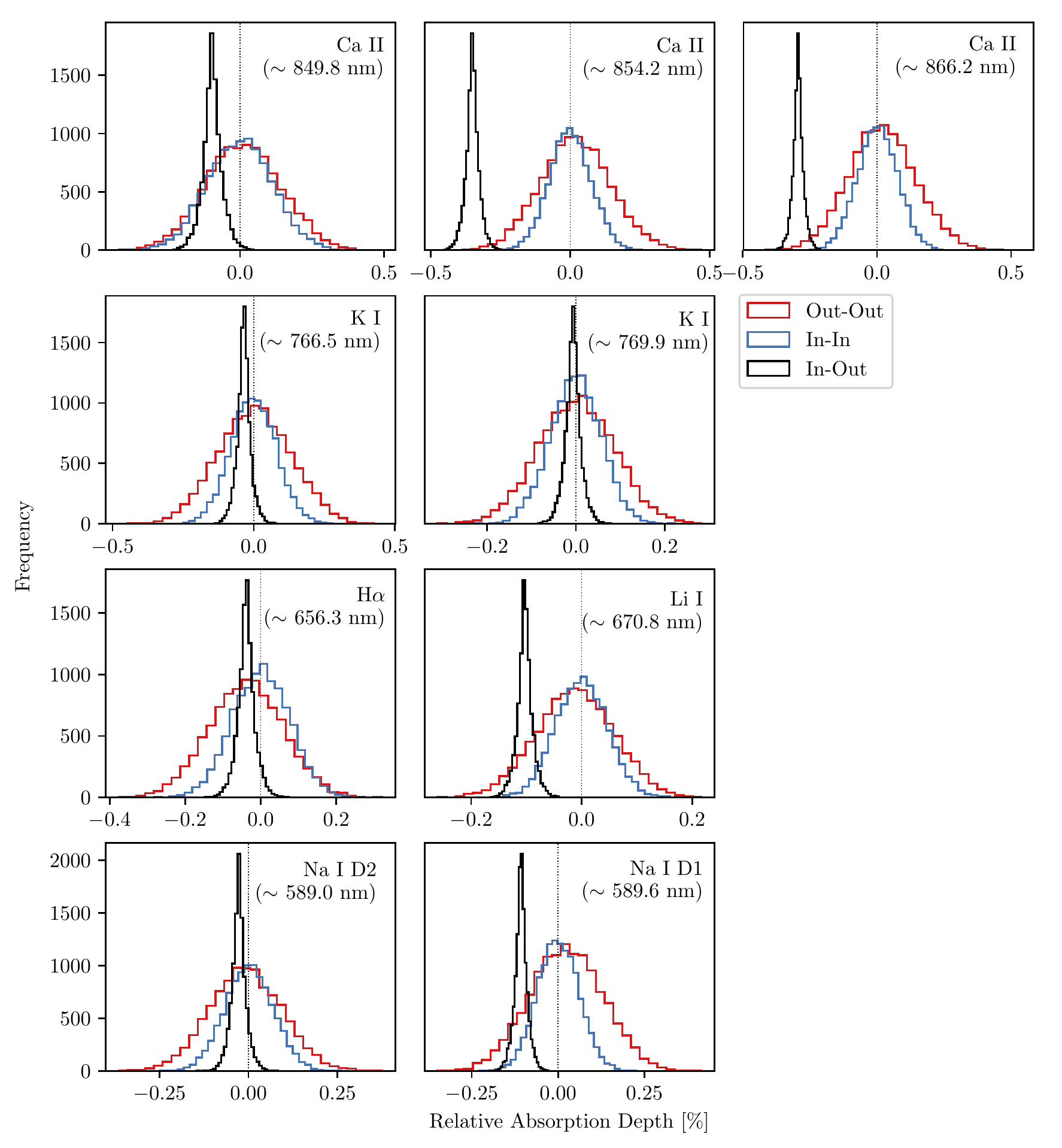}
    \caption{The results of the EMC analysis described in Appendix \ref{app:emc} for each line investigated in this Letter. For each subplot, the red line represents the out-out scenario, the blue line represents the in-in scenario, and the black line represents the in-out scenario. Note the differing x- and y-scales for each subplot. As expected, the in-in and out-out scenarios tend to be centered around zero, while the in-out scenarios for detected line species are centered around negative absorption depths. The differing widths of the distributions are due to the fact that different numbers of spectra are available for different scenarios, and do not have an impact on the analysis.}
    \label{fig:emc}
\end{figure}

The results for all lines investigated in this work are presented in Fig. \ref{fig:emc}. As can be seen in the figure, the out-out and in-in scenarios are centered around zero, while the in-out scenarios for detected line species are centered around negative absorption depths.

In the case of H$\alpha$, the in-out scenarios is slightly offset from zero, which does indicate a potential detection. However, we note that the out-out and in-in scenarios are also slightly offset from zero; while the widths of these distributions (i.e., the errors) are large enough to place the distributions within error of zero absorption, this could potentially indicate that residual stellar H$\alpha$ absorption is impacting our results. We thus do not classify this or the transmission spectrum presented in Fig. \ref{fig:ha-li} as a detection.

In the case of Li~I, the in-out scenario shows a considerable negative offset, while the in-in and out-out scenarios do not. While this would indicate a detection, the transmission spectrum presented in Fig. \ref{fig:ha-li} contains a number of dips/peaks adjacent to the absorption line profile that suggest that noise may be impacting our results. As in Appendix \ref{app:ha-li}, we thus classify this as a tentative detection warranting further investigation.

Finally, in the case of K~I, the in-out scenario of the line at $\sim$ 769.9 nm is only slightly offset from zero while the line at $\sim$ 766.5 nm is more consistent with a negative absorption depth. While the transmission spectra presented in Fig. \ref{fig:k} do indicate a potential detection, the results of the EMC analysis are inconclusive and suggest that these absorption features may in fact be caused by noise or other systematic effects in the data. We thus classify K~I as a tentative detection as well, but caution that systematic effects may be impacting our results.

\end{appendix}

\bibliography{references}{}

\begin{thebibliography}{}
\expandafter\ifx\csname natexlab\endcsname\relax\def\natexlab#1{#1}\fi
\providecommand{\url}[1]{\href{#1}{#1}}
\providecommand{\dodoi}[1]{doi:~\href{http://doi.org/#1}{\nolinkurl{#1}}}
\providecommand{\doeprint}[1]{\href{http://ascl.net/#1}{\nolinkurl{http://ascl.net/#1}}}
\providecommand{\doarXiv}[1]{\href{https://arxiv.org/abs/#1}{\nolinkurl{https://arxiv.org/abs/#1}}}

\bibitem[{{Allart} {et~al.}(2017){Allart}, {Lovis}, {Pino}, {Wyttenbach},
  {Ehrenreich}, \& {Pepe}}]{Allart17}
{Allart}, R., {Lovis}, C., {Pino}, L., {et~al.} 2017, \aap, 606, A144,
  \dodoi{10.1051/0004-6361/201730814}

\bibitem[{{Allart} {et~al.}(2018){Allart}, {Bourrier}, {Lovis}, {Ehrenreich},
  {Spake}, {Wyttenbach}, {Pino}, {Pepe}, {Sing}, \& {Lecavelier des
  Etangs}}]{Allart18}
{Allart}, R., {Bourrier}, V., {Lovis}, C., {et~al.} 2018, Science, 362, 1384,
  \dodoi{10.1126/science.aat5879}

\bibitem[{{Allart} {et~al.}(2020){Allart}, {Pino}, {Lovis}, {Sousa},
  {Casasayas-Barris}, {Zapatero Osorio}, {Cretignier}, {Palle}, {Pepe},
  {Cristiani}, {Rebolo}, {Santos}, {Borsa}, {Bourrier}, {Demangeon},
  {Ehrenreich}, {Lavie}, {Lendl}, {Lillo-Box}, {Micela}, {Oshagh}, {Sozzetti},
  {Tabernero}, {Adibekyan}, {Allende Prieto}, {Alibert}, {Amate}, {Benz},
  {Bouchy}, {Cabral}, {Dekker}, {D'Odorico}, {Di Marcantonio}, {Dumusque},
  {Figueira}, {Genova Santos}, {Gonz{\'a}lez Hern{\'a}ndez}, {Lo Curto},
  {Manescau}, {Martins}, {M{\'e}gevand}, {Mehner}, {Molaro}, {Nunes},
  {Poretti}, {Riva}, {Su{\'a}rez Mascare{\~n}o}, {Udry}, \& {Zerbi}}]{Allart20}
{Allart}, R., {Pino}, L., {Lovis}, C., {et~al.} 2020, \aap, 644, A155,
  \dodoi{10.1051/0004-6361/202039234}

\bibitem[{{Arcangeli} {et~al.}(2018){Arcangeli}, {D{\'e}sert}, {Line}, {Bean},
  {Parmentier}, {Stevenson}, {Kreidberg}, {Fortney}, {Mansfield}, \&
  {Showman}}]{Arcangeli18}
{Arcangeli}, J., {D{\'e}sert}, J.-M., {Line}, M.~R., {et~al.} 2018, \apjl, 855,
  L30, \dodoi{10.3847/2041-8213/aab272}

\bibitem[{{Bell} \& {Cowan}(2018)}]{Bell18}
{Bell}, T.~J., \& {Cowan}, N.~B. 2018, \apjl, 857, L20,
  \dodoi{10.3847/2041-8213/aabcc8}

\bibitem[{{Brogi} {et~al.}(2016){Brogi}, {de Kok}, {Albrecht}, {Snellen},
  {Birkby}, \& {Schwarz}}]{Brogi16}
{Brogi}, M., {de Kok}, R.~J., {Albrecht}, S., {et~al.} 2016, \apj, 817, 106,
  \dodoi{10.3847/0004-637X/817/2/106}

\bibitem[{{Casasayas-Barris} {et~al.}(2019){Casasayas-Barris}, {Pall{\'e}},
  {Yan}, {Chen}, {Kohl}, {Stangret}, {Parviainen}, {Helling}, {Watanabe},
  {Czesla}, {Fukui}, {Monta{\~n}{\'e}s-Rodr{\'\i}guez}, {Nagel}, {Narita},
  {Nortmann}, {Nowak}, {Schmitt}, \& {Zapatero Osorio}}]{CB19}
{Casasayas-Barris}, N., {Pall{\'e}}, E., {Yan}, F., {et~al.} 2019, \aap, 628,
  A9, \dodoi{10.1051/0004-6361/201935623}

\bibitem[{{Casasayas-Barris} {et~al.}(2021){Casasayas-Barris}, {Orell-Miquel},
  {Stangret}, {Nortmann}, {Yan}, {Oshagh}, {Palle}, {Sanz-Forcada},
  {L{\'o}pez-Puertas}, {Nagel}, {Luque}, {Morello}, {Snellen}, {Zechmeister},
  {Quirrenbach}, {Caballero}, {Ribas}, {Reiners}, {Amado}, {Bergond}, {Czesla},
  {Henning}, {Khalafinejad}, {Molaverdikhani}, {Montes}, {Perger},
  {S{\'a}nchez-L{\'o}pez}, \& {Sedaghati}}]{CB21}
{Casasayas-Barris}, N., {Orell-Miquel}, J., {Stangret}, M., {et~al.} 2021,
  arXiv e-prints, arXiv:2109.00059.
\newblock \doarXiv{2109.00059}

\bibitem[{{Chene} {et~al.}(2014){Chene}, {Padzer}, {Barrick}, {Anthony},
  {Benedict}, {Duncan}, {Gigoux}, {Kleinman}, {Malo}, {Martioli}, {Moutou},
  {Placco}, {Reshetovand}, {Rhee}, {Roth}, {Schiavon}, {Tollestrup},
  {Vermeulen}, {White}, \& {Wooff}}]{GRACES}
{Chene}, A.-N., {Padzer}, J., {Barrick}, G., {et~al.} 2014, in Society of
  Photo-Optical Instrumentation Engineers (SPIE) Conference Series, Vol. 9151,
  Advances in Optical and Mechanical Technologies for Telescopes and
  Instrumentation, ed. R.~{Navarro}, C.~R. {Cunningham}, \& A.~A. {Barto},
  915147, \dodoi{10.1117/12.2057417}

\bibitem[{{Deibert} {et~al.}(2019){Deibert}, {de Mooij}, {Jayawardhana},
  {Fortney}, {Brogi}, {Rustamkulov}, \& {Tamura}}]{Deibert19}
{Deibert}, E.~K., {de Mooij}, E. J.~W., {Jayawardhana}, R., {et~al.} 2019, \aj,
  157, 58, \dodoi{10.3847/1538-3881/aaf56b}

\bibitem[{{Ehrenreich} {et~al.}(2020){Ehrenreich}, {Lovis}, {Allart}, {Zapatero
  Osorio}, {Pepe}, {Cristiani}, {Rebolo}, {Santos}, {Borsa}, {Demangeon},
  {Dumusque}, {Gonz{\'a}lez Hern{\'a}ndez}, {Casasayas-Barris},
  {S{\'e}gransan}, {Sousa}, {Abreu}, {Adibekyan}, {Affolter}, {Allende Prieto},
  {Alibert}, {Aliverti}, {Alves}, {Amate}, {Avila}, {Baldini}, {Bandy}, {Benz},
  {Bianco}, {Bolmont}, {Bouchy}, {Bourrier}, {Broeg}, {Cabral}, {Calderone},
  {Pall{\'e}}, {Cegla}, {Cirami}, {Coelho}, {Conconi}, {Coretti}, {Cumani},
  {Cupani}, {Dekker}, {Delabre}, {Deiries}, {D'Odorico}, {Di Marcantonio},
  {Figueira}, {Fragoso}, {Genolet}, {Genoni}, {G{\'e}nova Santos}, {Hara},
  {Hughes}, {Iwert}, {Kerber}, {Knudstrup}, {Landoni}, {Lavie}, {Lizon},
  {Lendl}, {Lo Curto}, {Maire}, {Manescau}, {Martins}, {M{\'e}gevand},
  {Mehner}, {Micela}, {Modigliani}, {Molaro}, {Monteiro}, {Monteiro},
  {Moschetti}, {M{\"u}ller}, {Nunes}, {Oggioni}, {Oliveira}, {Pariani},
  {Pasquini}, {Poretti}, {Rasilla}, {Redaelli}, {Riva}, {Santana Tschudi},
  {Santin}, {Santos}, {Segovia Milla}, {Seidel}, {Sosnowska}, {Sozzetti},
  {Span{\`o}}, {Su{\'a}rez Mascare{\~n}o}, {Tabernero}, {Tenegi}, {Udry},
  {Zanutta}, \& {Zerbi}}]{Ehrenreich20}
{Ehrenreich}, D., {Lovis}, C., {Allart}, R., {et~al.} 2020, \nat, 580, 597,
  \dodoi{10.1038/s41586-020-2107-1}

\bibitem[{{Esteves} {et~al.}(2017){Esteves}, {de Mooij}, {Jayawardhana},
  {Watson}, \& {de Kok}}]{Esteves17}
{Esteves}, L.~J., {de Mooij}, E. J.~W., {Jayawardhana}, R., {Watson}, C., \&
  {de Kok}, R. 2017, \aj, 153, 268, \dodoi{10.3847/1538-3881/aa7133}

\bibitem[{{Ferland} {et~al.}(2017){Ferland}, {Chatzikos}, {Guzm{\'a}n},
  {Lykins}, {van Hoof}, {Williams}, {Abel}, {Badnell}, {Keenan}, {Porter}, \&
  {Stancil}}]{Ferland17}
{Ferland}, G.~J., {Chatzikos}, M., {Guzm{\'a}n}, F., {et~al.} 2017, \rmxaa, 53,
  385.
\newblock \doarXiv{1705.10877}

\bibitem[{{Fortney} {et~al.}(2008){Fortney}, {Marley}, {Saumon}, \&
  {Lodders}}]{Fortney08}
{Fortney}, J.~J., {Marley}, M.~S., {Saumon}, D., \& {Lodders}, K. 2008, \apj,
  683, 1104, \dodoi{10.1086/589942}

\bibitem[{{Fortney} {et~al.}(2020){Fortney}, {Visscher}, {Marley}, {Hood},
  {Line}, {Thorngren}, {Freedman}, \& {Lupu}}]{Fortney20}
{Fortney}, J.~J., {Visscher}, C., {Marley}, M.~S., {et~al.} 2020, \aj, 160,
  288, \dodoi{10.3847/1538-3881/abc5bd}

\bibitem[{{Fossati} {et~al.}(2021){Fossati}, {Young}, {Shulyak}, {Koskinen},
  {Huang}, {Cubillos}, {France}, \& {Sreejith}}]{Fossati21}
{Fossati}, L., {Young}, M.~E., {Shulyak}, D., {et~al.} 2021, arXiv e-prints,
  arXiv:2106.11263.
\newblock \doarXiv{2106.11263}

\bibitem[{{Fossati} {et~al.}(2020){Fossati}, {Shulyak}, {Sreejith}, {Koskinen},
  {Young}, {Cubillos}, {Lara}, {France}, {Rengel}, {Cauley}, {Turner},
  {Wyttenbach}, \& {Yan}}]{Fossati20}
{Fossati}, L., {Shulyak}, D., {Sreejith}, A.~G., {et~al.} 2020, \aap, 643,
  A131, \dodoi{10.1051/0004-6361/202039061}

\bibitem[{{Gu} {et~al.}(2003){Gu}, {Lin}, \& {Bodenheimer}}]{Gu03}
{Gu}, P.-G., {Lin}, D. N.~C., \& {Bodenheimer}, P.~H. 2003, \apj, 588, 509,
  \dodoi{10.1086/373920}

\bibitem[{{Harris} {et~al.}(2020){Harris}, {Jarrod Millman}, {van der Walt},
  {Gommers}, {Virtanen}, {Cournapeau}, {Wieser}, {Taylor}, {Berg}, {Smith},
  {Kern}, {Picus}, {Hoyer}, {van Kerkwijk}, {Brett}, {Haldane}, {Fern{\'a}ndez
  del R{\'\i}o}, {Wiebe}, {Peterson}, {G{\'e}rard-Marchant}, {Sheppard},
  {Reddy}, {Weckesser}, {Abbasi}, {Gohlke}, \& {Oliphant}}]{harris2020}
{Harris}, C.~R., {Jarrod Millman}, K., {van der Walt}, S.~J., {et~al.} 2020,
  arXiv e-prints, arXiv:2006.10256.
\newblock \doarXiv{2006.10256}

\bibitem[{Harrower \& Brewer(2003)}]{ColorBrewer}
Harrower, M., \& Brewer, C.~A. 2003, The Cartographic Journal, 40, 27,
  \dodoi{10.1179/000870403235002042}

\bibitem[{{Hawker} {et~al.}(2018){Hawker}, {Madhusudhan}, {Cabot}, \&
  {Gandhi}}]{Hawker18}
{Hawker}, G.~A., {Madhusudhan}, N., {Cabot}, S. H.~C., \& {Gandhi}, S. 2018,
  \apjl, 863, L11, \dodoi{10.3847/2041-8213/aac49d}

\bibitem[{{H{\o}g} {et~al.}(2000){H{\o}g}, {Fabricius}, {Makarov}, {Urban},
  {Corbin}, {Wycoff}, {Bastian}, {Schwekendiek}, \& {Wicenec}}]{SIMBAD}
{H{\o}g}, E., {Fabricius}, C., {Makarov}, V.~V., {et~al.} 2000, \aap, 355, L27

\bibitem[{{Hood} {et~al.}(2020){Hood}, {Fortney}, {Line}, {Martin}, {Morley},
  {Birkby}, {Rustamkulov}, {Lupu}, \& {Freedman}}]{Hood20}
{Hood}, C.~E., {Fortney}, J.~J., {Line}, M.~R., {et~al.} 2020, \aj, 160, 198,
  \dodoi{10.3847/1538-3881/abb46b}

\bibitem[{Hunter(2007)}]{Hunter:2007}
Hunter, J.~D. 2007, Computing in Science \& Engineering, 9, 90,
  \dodoi{10.1109/MCSE.2007.55}

\bibitem[{{Kesseli} \& {Snellen}(2021)}]{Kesseli21}
{Kesseli}, A.~Y., \& {Snellen}, I.~A.~G. 2021, \apjl, 908, L17,
  \dodoi{10.3847/2041-8213/abe047}

\bibitem[{{Kreidberg}(2015)}]{batman}
{Kreidberg}, L. 2015, \pasp, 127, 1161, \dodoi{10.1086/683602}

\bibitem[{{Lothringer} {et~al.}(2018){Lothringer}, {Barman}, \&
  {Koskinen}}]{Lothringer18}
{Lothringer}, J.~D., {Barman}, T., \& {Koskinen}, T. 2018, \apj, 866, 27,
  \dodoi{10.3847/1538-4357/aadd9e}

\bibitem[{{Manset} \& {Donati}(2003)}]{espadons}
{Manset}, N., \& {Donati}, J.-F. 2003, in Society of Photo-Optical
  Instrumentation Engineers (SPIE) Conference Series, Vol. 4843, Polarimetry in
  Astronomy, ed. S.~{Fineschi}, 425--436, \dodoi{10.1117/12.458230}

\bibitem[{{Martioli} {et~al.}(2012){Martioli}, {Teeple}, {Manset}, {Devost},
  {Withington}, {Venne}, \& {Tannock}}]{opera}
{Martioli}, E., {Teeple}, D., {Manset}, N., {et~al.} 2012, in Society of
  Photo-Optical Instrumentation Engineers (SPIE) Conference Series, Vol. 8451,
  Software and Cyberinfrastructure for Astronomy II, ed. N.~M. {Radziwill} \&
  G.~{Chiozzi}, 84512B, \dodoi{10.1117/12.926627}

\bibitem[{{Mayor} {et~al.}(2003){Mayor}, {Pepe}, {Queloz}, {Bouchy},
  {Rupprecht}, {Lo Curto}, {Avila}, {Benz}, {Bertaux}, {Bonfils}, {Dall},
  {Dekker}, {Delabre}, {Eckert}, {Fleury}, {Gilliotte}, {Gojak}, {Guzman},
  {Kohler}, {Lizon}, {Longinotti}, {Lovis}, {Megevand}, {Pasquini}, {Reyes},
  {Sivan}, {Sosnowska}, {Soto}, {Udry}, {van Kesteren}, {Weber}, \&
  {Weilenmann}}]{HARPS}
{Mayor}, M., {Pepe}, F., {Queloz}, D., {et~al.} 2003, The Messenger, 114, 20

\bibitem[{{Merritt} {et~al.}(2021){Merritt}, {Gibson}, {Nugroho}, {de Mooij},
  {Hooton}, {Lothringer}, {Matthews}, {Mikal-Evans}, {Nikolov}, {Sing}, \&
  {Watson}}]{Merritt21}
{Merritt}, S.~R., {Gibson}, N.~P., {Nugroho}, S.~K., {et~al.} 2021, arXiv
  e-prints, arXiv:2106.15394.
\newblock \doarXiv{2106.15394}

\bibitem[{{Nortmann} {et~al.}(2018){Nortmann}, {Pall{\'e}}, {Salz},
  {Sanz-Forcada}, {Nagel}, {Alonso-Floriano}, {Czesla}, {Yan}, {Chen},
  {Snellen}, {Zechmeister}, {Schmitt}, {L{\'o}pez-Puertas}, {Casasayas-Barris},
  {Bauer}, {Amado}, {Caballero}, {Dreizler}, {Henning}, {Lamp{\'o}n}, {Montes},
  {Molaverdikhani}, {Quirrenbach}, {Reiners}, {Ribas}, {S{\'a}nchez-L{\'o}pez},
  {Schneider}, \& {Zapatero Osorio}}]{Nortmann18}
{Nortmann}, L., {Pall{\'e}}, E., {Salz}, M., {et~al.} 2018, Science, 362, 1388,
  \dodoi{10.1126/science.aat5348}

\bibitem[{{Nugroho} {et~al.}(2020){Nugroho}, {Gibson}, {de Mooij}, {Watson},
  {Kawahara}, \& {Merritt}}]{Nugroho20}
{Nugroho}, S.~K., {Gibson}, N.~P., {de Mooij}, E. J.~W., {et~al.} 2020, \mnras,
  496, 504, \dodoi{10.1093/mnras/staa1459}

\bibitem[{{Parmentier} {et~al.}(2018){Parmentier}, {Line}, {Bean}, {Mansfield},
  {Kreidberg}, {Lupu}, {Visscher}, {D{\'e}sert}, {Fortney}, {Deleuil},
  {Arcangeli}, {Showman}, \& {Marley}}]{Parmentier18}
{Parmentier}, V., {Line}, M.~R., {Bean}, J.~L., {et~al.} 2018, \aap, 617, A110,
  \dodoi{10.1051/0004-6361/201833059}

\bibitem[{{Pepe} {et~al.}(2021){Pepe}, {Cristiani}, {Rebolo}, {Santos},
  {Dekker}, {Cabral}, {Di Marcantonio}, {Figueira}, {Lo Curto}, {Lovis},
  {Mayor}, {M{\'e}gevand}, {Molaro}, {Riva}, {Zapatero Osorio}, {Amate},
  {Manescau}, {Pasquini}, {Zerbi}, {Adibekyan}, {Abreu}, {Affolter}, {Alibert},
  {Aliverti}, {Allart}, {Allende Prieto}, {{\'A}lvarez}, {Alves}, {Avila},
  {Baldini}, {Bandy}, {Barros}, {Benz}, {Bianco}, {Borsa}, {Bourrier},
  {Bouchy}, {Broeg}, {Calderone}, {Cirami}, {Coelho}, {Conconi}, {Coretti},
  {Cumani}, {Cupani}, {D'Odorico}, {Damasso}, {Deiries}, {Delabre},
  {Demangeon}, {Dumusque}, {Ehrenreich}, {Faria}, {Fragoso}, {Genolet},
  {Genoni}, {G{\'e}nova Santos}, {Gonz{\'a}lez Hern{\'a}ndez}, {Hughes},
  {Iwert}, {Kerber}, {Knudstrup}, {Landoni}, {Lavie}, {Lillo-Box}, {Lizon},
  {Maire}, {Martins}, {Mehner}, {Micela}, {Modigliani}, {Monteiro}, {Monteiro},
  {Moschetti}, {Murphy}, {Nunes}, {Oggioni}, {Oliveira}, {Oshagh}, {Pall{\'e}},
  {Pariani}, {Poretti}, {Rasilla}, {Rebord{\~a}o}, {Redaelli}, {Santana
  Tschudi}, {Santin}, {Santos}, {S{\'e}gransan}, {Schmidt}, {Segovia},
  {Sosnowska}, {Sozzetti}, {Sousa}, {Span{\`o}}, {Su{\'a}rez Mascare{\~n}o},
  {Tabernero}, {Tenegi}, {Udry}, \& {Zanutta}}]{ESPRESSO}
{Pepe}, F., {Cristiani}, S., {Rebolo}, R., {et~al.} 2021, \aap, 645, A96,
  \dodoi{10.1051/0004-6361/202038306}

\bibitem[{{Perez} \& {Granger}(2007)}]{4160251}
{Perez}, F., \& {Granger}, B.~E. 2007, Computing in Science Engineering, 9, 21,
  \dodoi{10.1109/MCSE.2007.53}

\bibitem[{{Piskunov} \& {Valenti}(2017)}]{SME}
{Piskunov}, N., \& {Valenti}, J.~A. 2017, \aap, 597, A16,
  \dodoi{10.1051/0004-6361/201629124}

\bibitem[{{Price-Whelan} {et~al.}(2018){Price-Whelan}, {Sip{\H{o}}cz},
  {G{\"u}nther}, {Lim}, {Crawford}, {Conseil}, {Shupe}, {Craig}, {Dencheva},
  {Ginsburg}, {VanderPlas}, {Bradley}, {P{\'e}rez-Su{\'a}rez}, {de Val-Borro},
  {Paper Contributors}, {Aldcroft}, {Cruz}, {Robitaille}, {Tollerud},
  {Coordination Committee}, {Ardelean}, {Babej}, {Bach}, {Bachetti}, {Bakanov},
  {Bamford}, {Barentsen}, {Barmby}, {Baumbach}, {Berry}, {Biscani}, {Boquien},
  {Bostroem}, {Bouma}, {Brammer}, {Bray}, {Breytenbach}, {Buddelmeijer},
  {Burke}, {Calderone}, {Cano Rodr{\'\i}guez}, {Cara}, {Cardoso}, {Cheedella},
  {Copin}, {Corrales}, {Crichton}, {D{\textquoteright}Avella}, {Deil},
  {Depagne}, {Dietrich}, {Donath}, {Droettboom}, {Earl}, {Erben}, {Fabbro},
  {Ferreira}, {Finethy}, {Fox}, {Garrison}, {Gibbons}, {Goldstein}, {Gommers},
  {Greco}, {Greenfield}, {Groener}, {Grollier}, {Hagen}, {Hirst}, {Homeier},
  {Horton}, {Hosseinzadeh}, {Hu}, {Hunkeler}, {Ivezi{\'c}}, {Jain}, {Jenness},
  {Kanarek}, {Kendrew}, {Kern}, {Kerzendorf}, {Khvalko}, {King}, {Kirkby},
  {Kulkarni}, {Kumar}, {Lee}, {Lenz}, {Littlefair}, {Ma}, {Macleod},
  {Mastropietro}, {McCully}, {Montagnac}, {Morris}, {Mueller}, {Mumford},
  {Muna}, {Murphy}, {Nelson}, {Nguyen}, {Ninan}, {N{\"o}the}, {Ogaz}, {Oh},
  {Parejko}, {Parley}, {Pascual}, {Patil}, {Patil}, {Plunkett}, {Prochaska},
  {Rastogi}, {Reddy Janga}, {Sabater}, {Sakurikar}, {Seifert}, {Sherbert},
  {Sherwood-Taylor}, {Shih}, {Sick}, {Silbiger}, {Singanamalla}, {Singer},
  {Sladen}, {Sooley}, {Sornarajah}, {Streicher}, {Teuben}, {Thomas},
  {Tremblay}, {Turner}, {Terr{\'o}n}, {van Kerkwijk}, {de la Vega}, {Watkins},
  {Weaver}, {Whitmore}, {Woillez}, {Zabalza}, \& {Contributors}}]{astropy:2018}
{Price-Whelan}, A.~M., {Sip{\H{o}}cz}, B.~M., {G{\"u}nther}, H.~M., {et~al.}
  2018, \aj, 156, 123, \dodoi{10.3847/1538-3881/aabc4f}

\bibitem[{{Redfield} {et~al.}(2008){Redfield}, {Endl}, {Cochran}, \&
  {Koesterke}}]{Redfield08}
{Redfield}, S., {Endl}, M., {Cochran}, W.~D., \& {Koesterke}, L. 2008, \apjl,
  673, L87, \dodoi{10.1086/527475}

\bibitem[{{Seidel} {et~al.}(2019){Seidel}, {Ehrenreich}, {Wyttenbach},
  {Allart}, {Lendl}, {Pino}, {Bourrier}, {Cegla}, {Lovis}, {Barrado},
  {Bayliss}, {Astudillo-Defru}, {Deline}, {Fisher}, {Heng}, {Joseph}, {Lavie},
  {Melo}, {Pepe}, {S{\'e}gransan}, \& {Udry}}]{Seidel19}
{Seidel}, J.~V., {Ehrenreich}, D., {Wyttenbach}, A., {et~al.} 2019, \aap, 623,
  A166, \dodoi{10.1051/0004-6361/201834776}

\bibitem[{{Seidel} {et~al.}(2020){Seidel}, {Ehrenreich}, {Bourrier}, {Allart},
  {Attia}, {Hoeijmakers}, {Lendl}, {Linder}, {Wyttenbach}, {Astudillo-Defru},
  {Bayliss}, {Cegla}, {Heng}, {Lavie}, {Lovis}, {Melo}, {Pepe}, {dos Santos},
  {S{\'e}gransan}, \& {Udry}}]{Seidel20b}
{Seidel}, J.~V., {Ehrenreich}, D., {Bourrier}, V., {et~al.} 2020, \aap, 641,
  L7, \dodoi{10.1051/0004-6361/202038497}

\bibitem[{{Seidel} {et~al.}(2021){Seidel}, {Ehrenreich}, {Allart},
  {Hoeijmakers}, {Lovis}, {Bourrier}, {Pino}, {Wyttenbach}, {Adibekyan},
  {Alibert}, {Borsa}, {Casasayas-Barris}, {Cristiani}, {Demangeon}, {Di
  Marcantonio}, {Figueira}, {Gonz{\'a}lez Hern{\'a}ndez}, {Lillo-Box},
  {Martins}, {Mehner}, {Molaro}, {Nunes}, {Palle}, {Pepe}, {Santos}, {Sousa},
  {Sozzetti}, {Tabernero}, \& {Zapatero Osorio}}]{Seidel21}
{Seidel}, J.~V., {Ehrenreich}, D., {Allart}, A., {et~al.} 2021, arXiv e-prints,
  arXiv:2107.09530.
\newblock \doarXiv{2107.09530}

\bibitem[{{Sing} {et~al.}(2016){Sing}, {Fortney}, {Nikolov}, {Wakeford},
  {Kataria}, {Evans}, {Aigrain}, {Ballester}, {Burrows}, {Deming},
  {D{\'e}sert}, {Gibson}, {Henry}, {Huitson}, {Knutson}, {Lecavelier Des
  Etangs}, {Pont}, {Showman}, {Vidal-Madjar}, {Williamson}, \&
  {Wilson}}]{Sing2016}
{Sing}, D.~K., {Fortney}, J.~J., {Nikolov}, N., {et~al.} 2016, \nat, 529, 59,
  \dodoi{10.1038/nature16068}

\bibitem[{{Sing} {et~al.}(2019){Sing}, {Lavvas}, {Ballester}, {Lecavelier des
  Etangs}, {Marley}, {Nikolov}, {Ben-Jaffel}, {Bourrier}, {Buchhave}, {Deming},
  {Ehrenreich}, {Mikal-Evans}, {Kataria}, {Lewis}, {L{\'o}pez-Morales},
  {Garc{\'\i}a Mu{\~n}oz}, {Henry}, {Sanz-Forcada}, {Spake}, {Wakeford}, \&
  {PanCET Collaboration}}]{Sing19}
{Sing}, D.~K., {Lavvas}, P., {Ballester}, G.~E., {et~al.} 2019, \aj, 158, 91,
  \dodoi{10.3847/1538-3881/ab2986}

\bibitem[{{Snellen} {et~al.}(2008){Snellen}, {Albrecht}, {de Mooij}, \& {Le
  Poole}}]{Snellen08}
{Snellen}, I.~A.~G., {Albrecht}, S., {de Mooij}, E.~J.~W., \& {Le Poole}, R.~S.
  2008, \aap, 487, 357, \dodoi{10.1051/0004-6361:200809762}

\bibitem[{{Snellen} {et~al.}(2014){Snellen}, {Brandl}, {de Kok}, {Brogi},
  {Birkby}, \& {Schwarz}}]{Snellen14}
{Snellen}, I. A.~G., {Brandl}, B.~R., {de Kok}, R.~J., {et~al.} 2014, \nat,
  509, 63, \dodoi{10.1038/nature13253}

\bibitem[{{Snellen} {et~al.}(2010){Snellen}, {de Kok}, {de Mooij}, \&
  {Albrecht}}]{Snellen10}
{Snellen}, I.~A.~G., {de Kok}, R.~J., {de Mooij}, E.~J.~W., \& {Albrecht}, S.
  2010, \nat, 465, 1049, \dodoi{10.1038/nature09111}

\bibitem[{{Tabernero} {et~al.}(2021){Tabernero}, {Zapatero Osorio}, {Allart},
  {Borsa}, {Casasayas-Barris}, {Demangeon}, {Ehrenreich}, {Lillo-Box}, {Lovis},
  {Pall{\'e}}, {Sousa}, {Rebolo}, {Santos}, {Pepe}, {Cristiani}, {Adibekyan},
  {Allende Prieto}, {Alibert}, {Barros}, {Bouchy}, {Bourrier}, {D'Odorico},
  {Dumusque}, {Faria}, {Figueira}, {G{\'e}nova Santos}, {Gonz{\'a}lez
  Hern{\'a}ndez}, {Hojjatpanah}, {Lo Curto}, {Lavie}, {Martins}, {Martins},
  {Mehner}, {Micela}, {Molaro}, {Nunes}, {Poretti}, {Seidel}, {Sozzetti},
  {Su{\'a}rez Mascare{\~n}o}, {Udry}, {Aliverti}, {Affolter}, {Alves}, {Amate},
  {Avila}, {Bandy}, {Benz}, {Bianco}, {Broeg}, {Cabral}, {Conconi}, {Coelho},
  {Cumani}, {Deiries}, {Dekker}, {Delabre}, {Fragoso}, {Genoni}, {Genolet},
  {Hughes}, {Knudstrup}, {Kerber}, {Landoni}, {Lizon}, {Maire}, {Manescau}, {Di
  Marcantonio}, {M{\'e}gevand}, {Monteiro}, {Monteiro}, {Moschetti}, {Mueller},
  {Modigliani}, {Oggioni}, {Oliveira}, {Pariani}, {Pasquini}, {Rasilla},
  {Redaelli}, {Riva}, {Santana-Tschudi}, {Santin}, {Santos}, {Segovia},
  {Sosnowska}, {Span{\`o}}, {Tenegi}, {Iwert}, {Zanutta}, \&
  {Zerbi}}]{Tabernero20}
{Tabernero}, H.~M., {Zapatero Osorio}, M.~R., {Allart}, R., {et~al.} 2021,
  \aap, 646, A158, \dodoi{10.1051/0004-6361/202039511}

\bibitem[{{Tamuz} {et~al.}(2005){Tamuz}, {Mazeh}, \& {Zucker}}]{Tamuz2005}
{Tamuz}, O., {Mazeh}, T., \& {Zucker}, S. 2005, \mnras, 356, 1466,
  \dodoi{10.1111/j.1365-2966.2004.08585.x}

\bibitem[{{Turner} {et~al.}(2020){Turner}, {de Mooij}, {Jayawardhana}, {Young},
  {Fossati}, {Koskinen}, {Lothringer}, {Karjalainen}, \&
  {Karjalainen}}]{Turner20}
{Turner}, J.~D., {de Mooij}, E. J.~W., {Jayawardhana}, R., {et~al.} 2020,
  \apjl, 888, L13, \dodoi{10.3847/2041-8213/ab60a9}

\bibitem[{Virtanen {et~al.}(2020)Virtanen, Gommers, Oliphant, Haberland, Reddy,
  Cournapeau, Burovski, Peterson, Weckesser, Bright, {van der Walt}, Brett,
  Wilson, Millman, Mayorov, Nelson, Jones, Kern, Larson, Carey, Polat, Feng,
  Moore, {VanderPlas}, Laxalde, Perktold, Cimrman, Henriksen, Quintero, Harris,
  Archibald, Ribeiro, Pedregosa, {van Mulbregt}, \& {SciPy 1.0
  Contributors}}]{2020SciPy-NMeth}
Virtanen, P., Gommers, R., Oliphant, T.~E., {et~al.} 2020, Nature Methods, 17,
  261, \dodoi{10.1038/s41592-019-0686-2}

\bibitem[{{von Essen} {et~al.}(2020){von Essen}, {Mallonn}, {Hermansen},
  {Nixon}, {Madhusudhan}, {Kjeldsen}, \&
  {Tautvai{\v{s}}ien{\.{e}}}}]{vonEssen20}
{von Essen}, C., {Mallonn}, M., {Hermansen}, S., {et~al.} 2020, \aap, 637, A76,
  \dodoi{10.1051/0004-6361/201937169}

\bibitem[{{Wardenier} {et~al.}(2021){Wardenier}, {Parmentier}, {Lee}, {Line},
  \& {Gharib-Nezhad}}]{Wardenier21}
{Wardenier}, J.~P., {Parmentier}, V., {Lee}, E. K.~H., {Line}, M., \&
  {Gharib-Nezhad}, E. 2021, arXiv e-prints, arXiv:2105.11034.
\newblock \doarXiv{2105.11034}

\bibitem[{{West} {et~al.}(2016){West}, {Hellier}, {Almenara}, {Anderson},
  {Barros}, {Bouchy}, {Brown}, {Collier Cameron}, {Deleuil}, {Delrez}, {Doyle},
  {Faedi}, {Fumel}, {Gillon}, {G{\'o}mez Maqueo Chew}, {H{\'e}brard}, {Jehin},
  {Lendl}, {Maxted}, {Pepe}, {Pollacco}, {Queloz}, {S{\'e}gransan}, {Smalley},
  {Smith}, {Southworth}, {Triaud}, \& {Udry}}]{West16}
{West}, R.~G., {Hellier}, C., {Almenara}, J.~M., {et~al.} 2016, \aap, 585,
  A126, \dodoi{10.1051/0004-6361/201527276}

\bibitem[{{Wright} \& {Eastman}(2014)}]{barycorr}
{Wright}, J.~T., \& {Eastman}, J.~D. 2014, \pasp, 126, 838,
  \dodoi{10.1086/678541}

\bibitem[{{Wyttenbach} {et~al.}(2015){Wyttenbach}, {Ehrenreich}, {Lovis},
  {Udry}, \& {Pepe}}]{Wyttenbach15}
{Wyttenbach}, A., {Ehrenreich}, D., {Lovis}, C., {Udry}, S., \& {Pepe}, F.
  2015, \aap, 577, A62, \dodoi{10.1051/0004-6361/201525729}

\bibitem[{{Yan} {et~al.}(2019){Yan}, {Casasayas-Barris}, {Molaverdikhani},
  {Alonso-Floriano}, {Reiners}, {Pall{\'e}}, {Henning}, {Molli{\`e}re}, {Chen},
  {Nortmann}, {Snellen}, {Ribas}, {Quirrenbach}, {Caballero}, {Amado},
  {Azzaro}, {Bauer}, {Cort{\'e}s Contreras}, {Czesla}, {Khalafinejad}, {Lara},
  {L{\'o}pez-Puertas}, {Montes}, {Nagel}, {Oshagh}, {S{\'a}nchez-L{\'o}pez},
  {Stangret}, \& {Zechmeister}}]{Yan19}
{Yan}, F., {Casasayas-Barris}, N., {Molaverdikhani}, K., {et~al.} 2019, \aap,
  632, A69, \dodoi{10.1051/0004-6361/201936396}

\bibitem[{{Young} {et~al.}(2020){Young}, {Fossati}, {Koskinen}, {Salz},
  {Cubillos}, \& {France}}]{young2020}
{Young}, M.~E., {Fossati}, L., {Koskinen}, T.~T., {et~al.} 2020, \aap, 641,
  A47, \dodoi{10.1051/0004-6361/202037672}

\end{thebibliography}
\bibliographystyle{aasjournal}

\end{document}